\documentclass[aps,prd,reprint,superscriptaddress,preprintnumbers,floatfix,nofootinbib]{revtex4-1}
\usepackage{microtype}
\usepackage{comment}
\usepackage{bbm}

\usepackage{graphicx}
\usepackage[dvipsnames]{xcolor}
\usepackage{rotating}
\usepackage[caption=false,singlelinecheck=false]{subfig}
\usepackage{amsmath}
\usepackage{amssymb}
\usepackage{multirow}
\usepackage{amsfonts,dsfont}
\usepackage{bm}
\usepackage{physics}
\usepackage{mathtools}
\usepackage{slashed} 

\usepackage{adjustbox}
\usepackage{tikz}

\usepackage{booktabs}
\usepackage{array}
\usepackage{enumitem}
\usepackage[normalem]{ulem}

\usepackage{url}
\usepackage{xspace}
\usepackage{siunitx}
\usepackage{xfrac}
\usepackage{hyperref}
\usepackage[nameinlink]{cleveref}
\usepackage{appendix}

\usepackage{xifthen}
\usepackage{xcolor}
\hypersetup{colorlinks,	linkcolor={green!47!black},	citecolor={green!60!black},	urlcolor={green!55!black}}

\definecolor{ochre}{rgb}{0.8, 0.47, 0.13}
\definecolor{navyblue}{rgb}{0.0, 0.0, 0.5}
\definecolor{oucrimsonred}{rgb}{0.6, 0.0, 0.0}
\definecolor{nicegreen}{rgb}{0.31, 0.64, 0.38}

\usepackage{lmodern}     % set math font to Latin modern math
\usepackage[T1]{fontenc}

\DeclareSymbolFont{myletters}{OML}{ztmcm}{m}{it}
\DeclareMathSymbol{\uplambda}{\mathord}{myletters}{"15}

\newcommand{\imag}{\text{i}}

\newcolumntype{x}[1]{>{\centering\arraybackslash\hspace{0.cm}}p{#1}}

\DeclareSymbolFont{symbolsC}{U}{pxsyc}{m}{n}

\graphicspath{{figures/}}
\newcommand{\commentmute}[1]{} %%mute

\newcommand{\dSSB}{{\rm d}\chi{\rm SB}}

\newcommand{\Tcrit}{T_{\rm crit}}
\newcommand{\Tpre}{T_{\rm pre}}

\graphicspath{{../figures/},{figures/}}
\begin{document}
	\preprint{RIKEN-iTHEMS-Report-26}

	\title{Thermal precondensation in gauge-fermion theories}
	\author{\'Alvaro~Pastor-Guti\'errez}
	\affiliation{RIKEN Center for Interdisciplinary Theoretical and Mathematical Sciences (iTHEMS), RIKEN, Wako 351-0198, Japan}
	\author{Jan~M.~Pawlowski}
	\affiliation{Institut für Theoretische Physik, Universität Heidelberg, Philosophenweg 16, 69120 Heidelberg, Germany}
	\affiliation{ExtreMe Matter Institute EMMI, GSI Helmholtzzentrum für Schwerionenforschung mbH, Planckstr.\ 1, 64291 Darmstadt, Germany}
	\author{Franz~R.~Sattler}
	\affiliation{Fakult{\"a}t f{\"u}r Physik, Universit{\"a}t Bielefeld, D-33615 Bielefeld, Germany}
\begin{abstract}
Precondensation is a peculiar phenomenon in phase transitions, characterised by the occurrence of a condensate only over a finite range of length scales.
It is closely connected to the emergence of domains, pseudo-gapped phases and spatial inhomogeneities in equilibrium.
In this work, we show its occurrence in gauge-fermion theories in the chiral limit, close to the thermal chiral phase transition.
We further show that the precondensation regime becomes increasingly pronounced and extends over a wider temperature range as the number of fermion flavours is increased. 
We analyse the underlying dynamics which is shared by a broad class of fermionic systems, ranging from condensed matter to high-energy physics. Specifically, we discuss the potential relevance of this phenomenon for physics beyond the Standard~Model.
\end{abstract}
\maketitle
%

%%%%%%%%%%%%%%%%%%%%%%%%%%%%%%%%%%%%%%%%
\section{Introduction}
\label{sec:Introduction}

Precondensation is characterised by the occurrence of a condensate only on finite length scales, while the macroscopic occupation vanishes. Many systems exhibit this phenomenon when approaching a phase transition from the symmetric phase.
This peculiar behaviour leads to characteristic features shared with related phenomena, such as generic spatial modulations, moat regimes, inhomogeneous condensates, or domain structures.

Typically, precondensation occurs in mixed systems where the dynamics of some modes triggers spontaneous symmetry breaking (SSB), while that of others is symmetry-restoring. Each mode dominates in characteristic momentum regimes and in general, variations of external parameters such as temperature, density, or volume affect their dynamics in different ways. 
The resulting momentum-dependent shift in the relative importance of the competing modes can then induce a momentum-dependent condensate within specific parameter regimes. 
In the present work, we use temperature as the external control parameter and study thermal phase transitions in gauge-fermion systems, where this intertwined dynamics gives rise to a momentum-dependent condensate below the precondensation temperature $\Tpre$ and above the critical temperature $\Tcrit$.

Due to its generality, precondensation arises in a broad class of finite-temperature systems. 
In condensed matter physics, it occurs in several low-dimensional fermionic and bosonic systems, see e.g.~\cite{Tolosa-Simeon:2025fot, Hawashin:2024dpp}. 
In cold-atom setups, it has been studied along the BCS-BCE crossover in pseudo-gapped phases, see e.g.~\cite{Boettcher:2012cm, Boettcher:2012dh, Boettcher:2013kia}. 
In high-energy physics, precondensation of colour-superconducting diquarks has been observed in two-colour QCD at finite temperature and chemical potential~\cite{Khan:2015puu}.
Signatures of this behaviour, without being linked to precondensation, have also been reported for the chiral condensate in low-energy effective theories of QCD in the chiral limit, see e.g.~\cite{Berges:1997eu, Schaefer:2004en, Herbst:2013ail, Aoki:2014ova, Aoki:2015hsa, Aoki:2017rjl, Dupuis:2020fhh}. 

The momentum dependence of the condensate in the precondensation regime may be accompanied by inhomogeneous structures in position space. 
This connects the phenomenon to regimes with spatial modulations in physical QCD, such as the moat regime at intermediate densities~\cite{Fu:2019hdw, Pisarski:2021qof, Rennecke:2023xhc, Fu:2024rto, Pawlowski:2025jpg}, or inhomogeneous condensates at high baryon density, see e.g.~\cite{Deryagin:1992rw, Kojo:2009ha, Carignano:2010ac, Fukushima:2010bq, Fukushima:2013rx, Lee:2015bva, Buballa:2018hux, Pisarski:2020gkx, Pisarski:2020dnx, Motta:2023pks, Motta:2024agi, Motta:2024rvk, Pawlowski:2025jpg} and \cite{Buballa:2014tba} for a review.  

\begin{figure*}[t!]
	\centering
	\includegraphics[width=.975\textwidth]{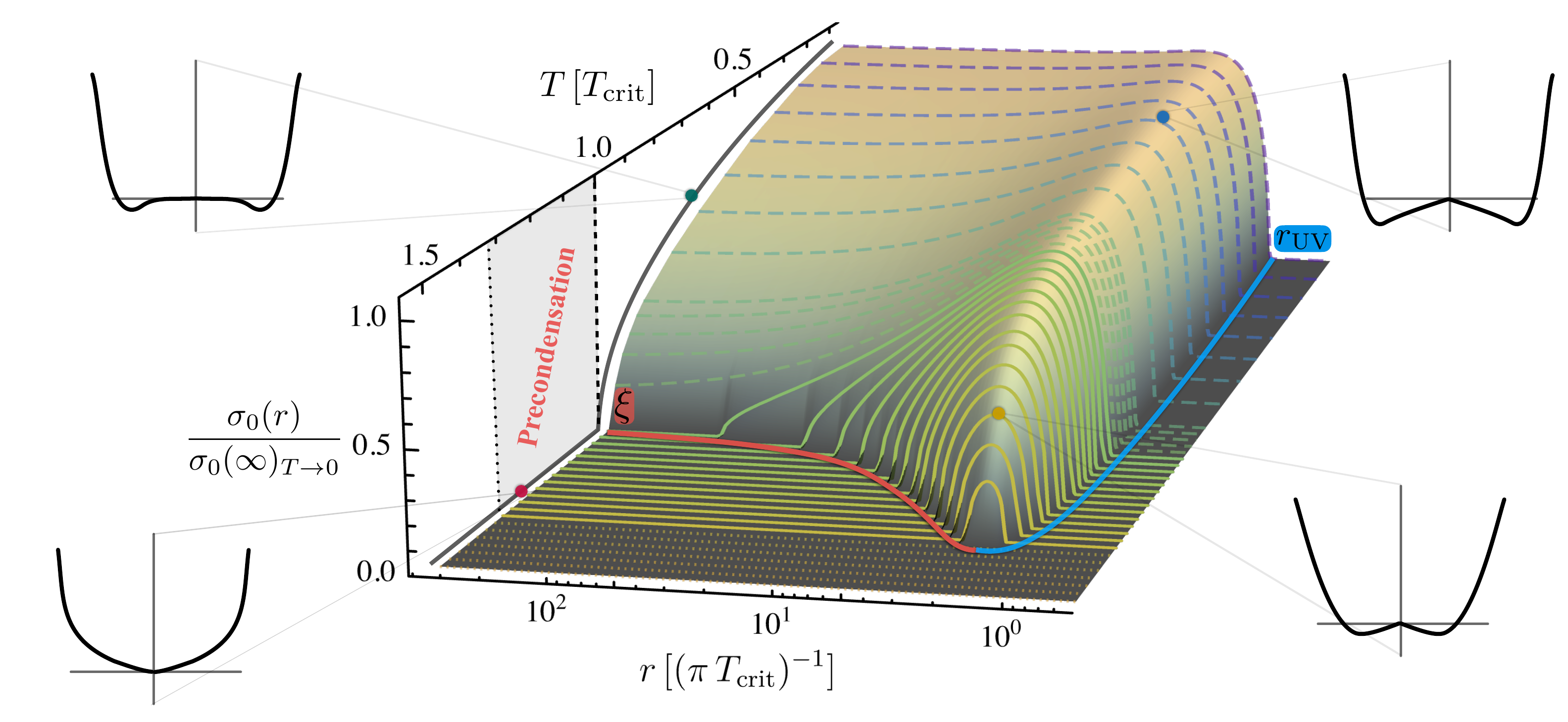}
    \caption{ Chiral order parameter $\sigma_0(r)$ of a QCD-like gauge-fermion theory with $N_c=3$ and $N_f=3$ chiral flavours as a function of the spatial separation $r$ (identified with the inverse fRG cutoff $k^{-1}$) and temperature, normalised to its value at zero temperature in the macroscopic limit ($\sigma_0(\infty)_{\,T\to0}$).
    The dotted lines correspond to the condensate at high temperatures, where the system remains in the symmetric phase at all length scales. Solid lines show the condensate in the precondensation regime, {$\Tcrit$~(vertical~dashed~line)~$<T<\Tpre$~(vertical~dotted line)}, where it is nonvanishing only over a finite range of length scales. 
    The dashed lines show temperatures in the broken phase, where a macroscopic condensate is obtained at large distances, $r\to\infty$.
    The blue line indicates the UV length scale ($r_{\rm UV}$), below which the condensate effectively vanishes. The red line marks the domain size ($\xi$) associated with precondensation, see \Cref{fig:cartoondomains}.
}
	\label{fig:OrderParameter}
\end{figure*}

In this work, we analyse the occurrence of precondensation in the novel context of QCD-like gauge-fermion theories in the chiral limit, following the suggestion thereof in~\cite{Goertz:2024dnz}. 
We show with a first-principles functional approach that these systems exhibit a precondensation regime alongside dynamical chiral symmetry breaking ($\dSSB$), see \Cref{fig:OrderParameter}. The infrared (IR) dynamics of massless Goldstone bosons is key to the underlying mechanism, and its enhancement with increasing flavour number $N_f$ indicates that precondensation becomes increasingly pronounced towards the conformal limit.
We support the significant rôle of the mechanism for the thermal properties of many-flavour theories with results for $N_f=2$, 3 and 4.
Furthermore, we also analyse the microscopic dynamics of precondensation, which allows us to dissect its necessary ingredients. These criteria are generic and do not only apply to the gauge-fermion theories studied here but also to condensed matter and cold-atom settings.

The gauge-fermion systems studied here naturally appear in new physics extensions of the Standard Model as candidates for addressing open problems and puzzles, see e.g.~\cite{Holthausen:2013ota, Kubo:2014ova, Iso:2017uuu, Reichert:2021cvs, Pasechnik:2023hwv, Goertz:2023nii, deBoer:2025vpx}. Specifically, they have attracted interest due to the possibility of displaying a first-order chiral phase transition, which could generate observable gravitational-wave signals in the early Universe. As we show, precondensation also introduces characteristic features that may lead to observable imprints, independent of the order of the transition. This is analogous to the moat regime and inhomogeneous condensation in QCD, which may be probed with Hanbury Brown-Twiss interferometry~\cite{Rennecke:2023xhc, Fukushima:2023tpv} through the formation of local clusters in the thermal medium. Such signatures may offer new avenues for testing physics beyond the Standard Model.

%%%%%%%%%%%%%%%%%%%%%%%%%%%%%%%%%%%%%%%%
\section{Thermal Precondensation}
\label{sec:Precondensation}

Precondensation occurs in a broad class of thermal systems in which a condensate forms only over a finite range of length scales and vanishes in the macroscopic limit. It occurs within a temperature window above the critical temperature and below the precondensation one, $\Tcrit \leq T \leq \Tpre$.
In \Cref{fig:OrderParameter}, we illustrate the occurrence of this phenomenon for a gauge-fermion theory in the chiral limit.
This constitutes one of the central new results of the present work and is discussed in detail in \Cref{sec:GaugeFermion}. Here, it serves to highlight the generic features of precondensation.

\Cref{fig:OrderParameter} displays the condensate $\sigma_0(r)$ as a function of the distance $r$ for different temperatures, normalised to its macroscopic value at $T=0$. This condensate is derived from the two-point correlation of the condensate field, averaged over spatial domains with radius $r$.

Above the precondensation regime, at $T > \Tpre$, the theory is in the symmetric phase.
For $T<\Tpre$, the non-trivial condensate emerges on a finite range of length scales, ${\sigma_0(r)\neq 0}$ for $r_{\textrm{UV}} < r < \xi$,
and vanishes in the macroscopic limit, ${\sigma_0(r\to\infty)=0}$.
The UV ($r_\textrm{UV}$) and IR ($\xi$) length scales are respectively marked by a blue and red line in \Cref{fig:OrderParameter}. The latter defines a correlation length of the precondensation phenomenon.    
For even smaller temperatures, $T\leq\Tcrit$, a macroscopic condensate emerges, i.e.~$\sigma(r\!\to\!\infty)\neq 0$.\\
Altogether, three different regimes can be identified:
\begin{align*}
    \begin{array}{rccl} 
	\textrm{(i)} & T< \Tcrit &:& \textit{Broken phase}\\[2ex]
	\textrm{(ii)} & \Tcrit< T < \Tpre &:&  \textit{Precondensation phase}\\[2ex]
	\textrm{(iii)} & T> \Tpre & :&  \textit{Symmetric phase}
\end{array}
\end{align*}
The occurrence of precondensation in the thermal evolution of a system may lead to spatial modulations, inhomogeneities or other modifications of the condensate background. In any case, its presence certainly leaves its trace in observables that resolve the spatial or spatial momentum dependence of the system.

%%%%%%%%%%%%%%%%%%%%%%%%%%%%%%%%%%%%%%%%
\subsection{Inhomogeneous condensates} 
\label{sec:PrecondensationBasics}

The ordered phase (i) at ${T<\Tcrit}$ is signalled by the emergence of a homogeneous condensate, 
\begin{align} 
	\sigma_0^2= \lim _{{\cal V}\to \infty } \frac{1}{{\cal V}^2} \int_{\cal V}  d^3 x\,d^3 y\,\langle \hat\sigma(x)\,\hat \sigma(y)\rangle \,,  
	\label{eq:Condensate2}
\end{align}
where ${\cal V}$ denotes the spatial volume of the system and $\hat\sigma$ is the fundamental field associated with the order-parameter. We have set $x_0=y_0$ and restricted ourselves to three spatial dimensions. 
A non-vanishing $\sigma_0^2 \neq 0$ in \Cref{eq:Condensate2} implies a macroscopic occupation of $\hat\sigma$. 

An equivalent definition of the homogeneous macroscopic condensate can also be obtained from the expectation value of the single operator $\hat\sigma$,
\begin{align} 
 	\sigma_0=\lim _{{\cal V}\to \infty } \frac{1}{\cal V} \int d^3 x\, \sigma_0(x)\,,\quad \sigma_0(x)=\langle \hat\sigma(x)\rangle \,.
 	\label{eq:Condensate}
\end{align}
The spatially dependent field expectation value $\sigma_0(x)$ can be thus decomposed into the inhomogeneous form
\begin{align} 
	&\sigma_0(x)=\sigma_0 + \Delta\sigma_0(x)\,,&&{\rm with}&&\int d^3 x \,\Delta\sigma_0(x)=0\,,
\label{eq:sigmax} 
\end{align} 
where $\Delta\sigma_0(x)$ carries the spatial modulation. This general form makes it explicit that the system can exhibit order at a finite correlation length $\xi$, even if $\sigma_0 = 0$ as $\Delta\sigma_0(x)$ could be non-vanishing.

Adapting the definition \labelcref{eq:Condensate2}, such a localised order can be measured using the equal-time two-point function averaged over a ball of radius $r$, centred around $\boldsymbol{x}$, 
\begin{align} 
	\sigma_0^2(\boldsymbol{x};r) = \frac{1}{{\cal V}_r}\int  d^3 y \,\langle \hat\sigma(x)\,\hat\sigma(y)\rangle\, \theta\left(r^2-(\boldsymbol{x}-\boldsymbol{y})^2\right)\,,
	\label{eq:Condensate-r}
\end{align} 
where ${\cal V}_r$ is the spatial volume of the ball. While 
for $r \approx \xi$ \labelcref{eq:Condensate-r} reflects the local order, we recover the homogeneous part $\sigma_0^2({x};r) \to \sigma_0^2$ in the limit $r\to \infty$.
This shows how by varying $r$ we can resolve the theory or rather the condensate at different length scales.

The Fourier transformation of $\sigma_0^2(\boldsymbol{x};r)$ reads
\begin{align} 
	\sigma_0^2(\boldsymbol{p};r)=  &\,\frac{1}{{\cal V}_r} \int d^3 q \,\left\langle \hat\sigma(\boldsymbol{p}-\boldsymbol{q})\hat\sigma(\boldsymbol{q})\right\rangle\,\delta_r(\boldsymbol{q} )\,,
		\label{eq:Condensate-rp}
\end{align} 
with the volume-regularised ${\delta\text{-function}}$
\begin{align} 
    \delta_r(\boldsymbol{q} )
    &=\int \frac{d^3 z}{(2\pi)^3}\,   e^{\imag \boldsymbol{q} \boldsymbol{z}}\,\theta\left(r^2-\boldsymbol{z}^2\right)\,,
\label{eq:delta-r}
\end{align} 
which removes all wave lengths longer than $r$, as they do not fit into the finite volume $(\boldsymbol{x}-\boldsymbol{y})^2\leq r^2$. The~corresponding momentum representation of~\labelcref{eq:sigmax} is once again a combination of the macroscopic zero mode $\sigma_0$ and a momentum-dependent `tail' $\Delta \sigma_0(\boldsymbol{p})$, the Fourier transform of $\Delta\sigma_0(\boldsymbol{x})$. We find 
\begin{align} 
	\sigma_0(\boldsymbol{p} ) = \sigma_0\,  (2\pi)^3 \delta(\boldsymbol{p} ) + \Delta\sigma_0(\boldsymbol{p} )\,,
\label{eq:sigmap}
\end{align}  
with $\Delta\sigma_0(0)=0$, which follows from the vanishing spatial integral in \labelcref{eq:sigmax}. Assuming the dominance of the disconnected terms in the two-point function, we can approximate the two-point function with 
${\langle \hat\sigma(\boldsymbol{p}-\boldsymbol{q})\hat\sigma(\boldsymbol{q})\rangle \approx\sigma_0(\boldsymbol{p}-\boldsymbol{q})\,\sigma_0(\boldsymbol{q})}$. Using the parametrisation \labelcref{eq:sigmap} in \labelcref{eq:Condensate-rp} leads us to the complete form
\begin{align} \nonumber 
	\sigma_0^2(\boldsymbol{p};r)=  &\,\sigma_0^2 (2\pi)^3 \delta(\boldsymbol{p})\\[1ex] \nonumber 
	& + \sigma_0 \,\Delta\sigma_0(\boldsymbol{p}) \left( 1 +  \, \frac{(2\pi)^3 }{{\cal V}_r} \delta_r(\boldsymbol{p}) \right) \\[1ex] 
	&+\frac{1}{{\cal V}_r} \int d^3 q \,\Delta\sigma_0(\boldsymbol{p}-\boldsymbol{q})\Delta\sigma_0(\boldsymbol{q})\,\delta_r(\boldsymbol{q} )\,.
	\label{eq:Condensate-rpApprox}
\end{align} 
In the macroscopic limit,
\begin{align} 
	\sigma_0^2(\boldsymbol{p}; r\to \infty)=  \sigma_0^2 (2\pi)^3 \delta(\boldsymbol{p})+ \sigma_0 \,\Delta\sigma_0(\boldsymbol{p}) \,, 
	\label{eq:Condensate-rInfty}
\end{align} 
which in position space corresponds to $\sigma^2_0(\boldsymbol{x})= \sigma_0^2 + \sigma_0 \Delta\sigma_0(\boldsymbol{x})$ as in \eqref{eq:sigmax}. While the latter term may be dropped if the modulation has a small amplitude relative to $\sigma_0$, in the absence of such macroscopic condensate the modulation 
$\Delta\sigma_0(\boldsymbol{x})$ in the last term of~\eqref{eq:Condensate-rpApprox} becomes the dominant contribution. 
In this case, the $r$-dependence of $\sigma_0^2(\boldsymbol{p},r)$ exposes the structure encoded in this modulation: \\[-2ex]

\begin{figure}[t!]
	\centering
	\includegraphics[width=.9\columnwidth]{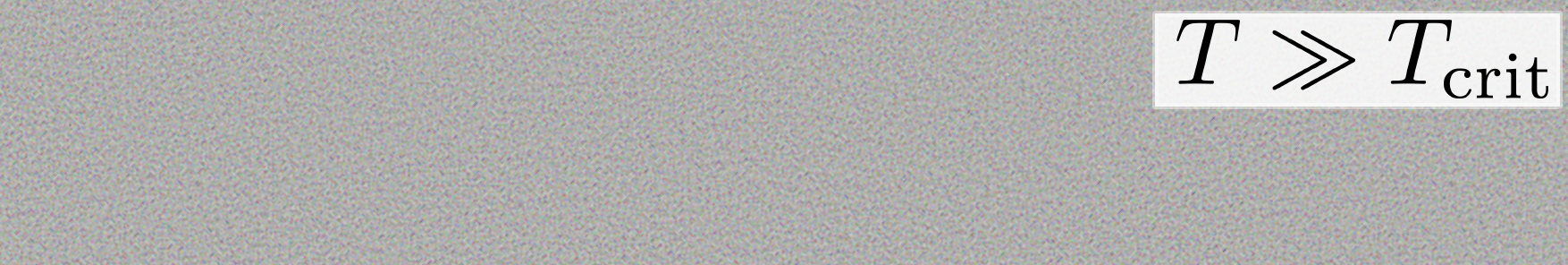}
	\vspace{.1cm}
	\includegraphics[width=.9\columnwidth]{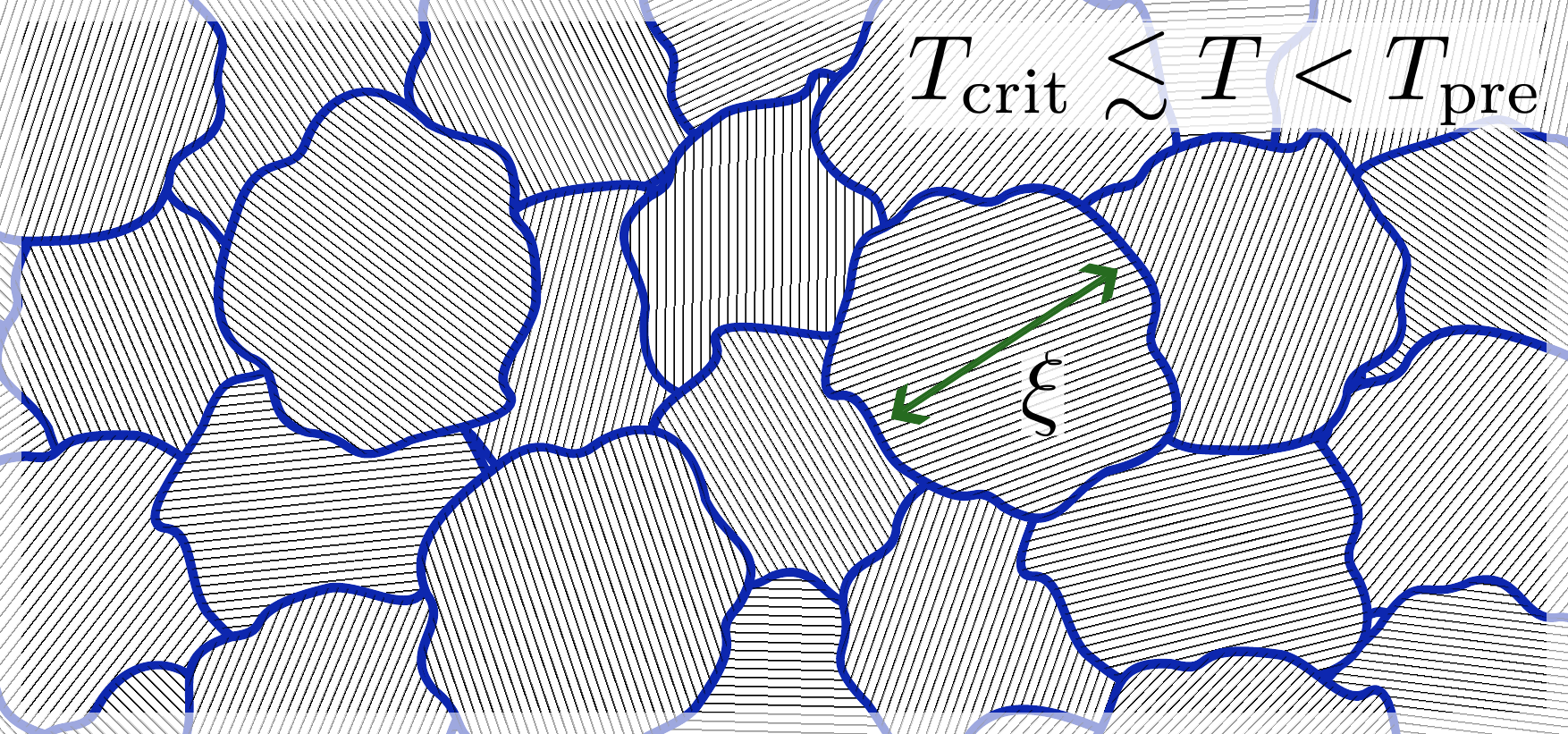}
	\vspace{.1cm}
	\includegraphics[width=.9\columnwidth]{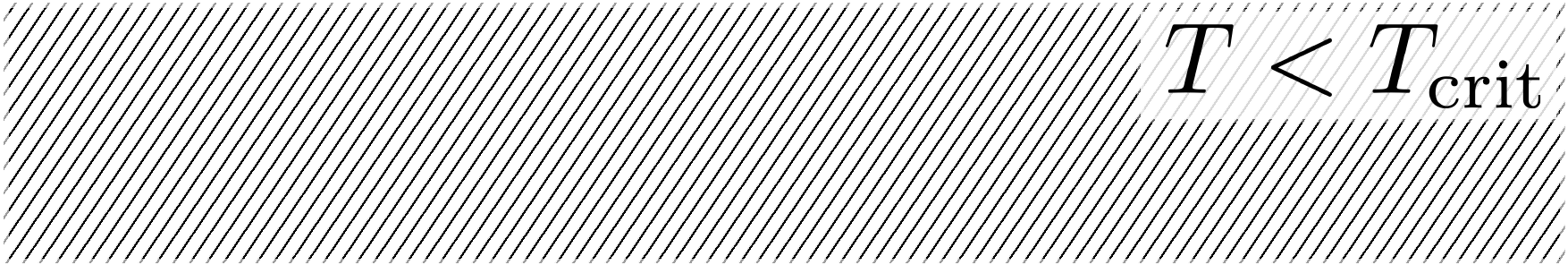}
	\caption{Cartoon of the thermal evolution of a system with precondensation at a fixed length scale $r\sim (\pi \Tcrit)^{-1}$.\\
    Top: symmetric phase at high temperatures where only fluctuations are present.\\
	Middle: precondensation phase at $\Tcrit < T < \Tpre$, where a local condensate exists on scales of order $\xi$, while $\sigma_0 = 0$ due to cancellations among domains at large distances.\\
	Bottom: broken phase below $\Tcrit$, where the domains align and a homogeneous macroscopic condensate forms.}
	\label{fig:cartoondomains}
\end{figure}
For $r \gg \xi$ the contribution of the modulation vanishes and the macroscopic limit is recovered, while for $r \lesssim \xi$ it remains finite and resolves the intrinsic scale of the modulation. For even shorter distances, $r < r_{\rm UV}$, physics is dominated by the (symmetric) UV fluctuations and the condensate consequently vanishes. 
Altogether, this yields
\begin{align}
	\sigma_0^2(\boldsymbol{p};r) = 
	\begin{cases}
		0, & r < r_{\rm UV}\,, \\[2ex]
		\neq 0, & r_{\rm UV} < r < \xi\,, \\[2ex]
		0, & r > \xi\,. 
	\end{cases}
	\label{eq:3Regimes-sigma}
\end{align}
This characteristic profile of a precondensing system is exemplified with the picture of domain formation, where the condensate locally aligns along different directions in theory (vacuum) space.  
It is analogous to Weiss domains in ferromagnets, where the magnetisation is locally nonzero but averages to zero at sufficiently large distances due to cancellations among differently oriented domains, see \Cref{fig:cartoondomains}.

Furthermore, at finite temperatures the dynamics dominating at different length or momentum scales is reweighted. This can trigger or remove spatial modulations for a given temperature. The different regimes can be characterised with the condensate profiles as in \Cref{fig:OrderParameter}, 
\begin{itemize}
	\item[(i)] \textit{Symmetric regime:} $\sigma_0^2(\boldsymbol{p};r) \equiv 0$.
	\item[(ii)] \textit{Precondensation regime:}  $\sigma_0^2(\boldsymbol{p};r) \neq 0$ and $\sigma_0 = 0$. 
	\item[(iii)] \textit{Broken phase:} $\sigma_0 \neq 0$.
\end{itemize}
This concludes our general discussion of the precondensation phenomenon.

%%%%%%%%%%%%%%%%%%%%%%%%%%%%%%%%%%%%%%%%
\subsection{Local order and the renormalisation group} 
\label{sec:fRG}

In the present work, we will access the quantum and thermal dynamics using the functional renormalisation group (fRG) approach. It is based on the Wilsonian idea of including quantum and thermal fluctuations step by step, which is achieved by introducing an IR regulator function $R_k(\boldsymbol{p})$ that suppresses modes with momenta~$\boldsymbol{p}^2 \lesssim k^2$,
\begin{align} 
	\int \!\prod_p d\phi(p)\to 		\int\! \prod_p d\phi(p)\,e^{-\frac12 \int  d^4 q \,\hat\sigma(p)\,R_k(\boldsymbol{p})\,\hat\sigma(-p)} \,. 
	\label{eq:RegPImeasure}
\end{align} 
This regulator is then removed by taking the limit ${k\to0}$, giving rise to a flow that progressively includes all fluctuations of the given theory.

\Cref{eq:RegPImeasure} introduces a direct analogue of \eqref{eq:Condensate-rp}: the spatial regulator function \eqref{eq:delta-r} suppresses all modes larger than $r$, i.e. of momenta $|\boldsymbol{p}| < 1/r$. On the other hand, the regulator $R_k(\boldsymbol{p})$ suppresses all modes below the IR cutoff scale $|\boldsymbol{p}| < k$ on the level of microscopic field operators.

With this mapping in mind, we present in the following Section our results, derived in this non-perturbative functional approach. 
We provide additional details on the explicit mapping between the spatial picture and momentum cutoff to show the direct connection $k\sim 1/r$, which can be found in \Cref{app:fRG}.

%%%%%%%%%%%%%%%%%%%%%%%%%%%%%%%%%%
\section{Precondensation~in~gauge-fermion~theories} 
\label{sec:GaugeFermion}

The strongly-correlated nature of QCD-like theories can only be accessed with non-perturbative approaches. In~this work, we use a first-principles functional RG approach which has been quantitatively developed in this context over the past two decades, and builds in particular on the recent advances of~\cite{Ihssen:2024miv, Goertz:2024dnz, Pawlowski:2025jpg}.  

%%%%%%%%%%%%%%%%%%%%%%%%%%%%%%%%%%
\subsection{Gauge-fermion theories in the chiral limit} 
\label{sec:GaugeFermionIntro}
The theories targeted in this work are encompassed by the classical action for gauge-fermion systems
\begin{align}
	S = \int_x \bigg\{ \,\frac14 F_{\mu\nu}^a F_{\mu\nu}^a+\bar \psi\, \gamma_\mu {D}_\mu\psi \,\bigg\} \,.
	\label{eq:action gauge-fermion}
\end{align}
This consists of a general Yang-Mills part with the field-strength tensor $F_{\mu\nu}^a$ and Dirac fermions $\psi$, which couple to the gauge sector via the covariant derivative $D_\mu=\partial_\mu -i g_s A_\mu$. We consider the local gauge symmetry SU($N_c$) with $N_c=3$ and fermions transforming under its fundamental representation, in analogy with physical QCD. Additionally, we choose the fermion sector to contain $N_f$ identical and massless flavours rendering the classical action $S$ symmetric under the exact global flavour symmetry given by~
$\textrm{SU}(N_f)_\textrm{L}\times\textrm{SU}(N_f)_\textrm{R}\times \textrm{U}(1)_\textrm{L}\times \textrm{U}(1)_\textrm{R}$.

Gauge-fermion theories described by \eqref{eq:action gauge-fermion} and with ${N_f \lesssim 11/2N_c}$ but non-conformal, are known to display two intertwined IR phenomena: colour confinement, associated with the generation of a mass gap in the gauge sector, and dynamical chiral symmetry breaking ($\dSSB$), characterised by the emergence of a chiral condensate. 
While the former occurs in the pure-gauge sector, the latter is encoded in solely fermionic $n$-point functions but driven by gauge-fermion interactions. The precondensation phenomenon is associated directly with the fermionic $\dSSB$ dynamics.

In the chiral limit, the flavour symmetry is exact, and the system undergoes a phase transition towards the chirally broken regime, where the global symmetry breaks down into a subgroup with pattern
\begin{align}
    \textrm{SU}(N_f)_{\rm L}\times\textrm{SU}(N_f)_{\rm R} \to \textrm{SU}(N_f)_{\rm V}\,.\notag
\end{align}
While in the non-chiral limit the positivity of the Euclidean action ensures that SU($N_f$)$_{\rm V}$~\cite{Vafa:1983tf} remains unbroken, such a proof does not yet exist for the chiral limit. Furthermore, we assume the same scenario and hence that the breaking is characterised by a condensate in the single radial $\sigma$-mode, accompanied by $N_f^2-1$ exact Nambu-Goldstone bosons, namely the pseudo-scalar ${\boldsymbol{\pi}\textrm{-modes}}$. The remaining scalar-pseudoscalar mesons (the $\eta$ and $a$'s) are lifted in mass due to the anomalous axial symmetry breaking induced by the axial anomaly. 
%~\cite{Adler:1969gk, Bell:1969ts, tHooft:1976rip, Veneziano:1979ec}. 
Although this last effect is decisive in the few-flavour limit, where the non-degeneracy of modes is significant, it becomes suppressed in the many-flavour regime ($N_f\gtrsim4$), since the associated ’t~Hooft operator increases in dimension.

Although both confinement and $\dSSB$ are highly non-trivial effects, they can be analysed at the level of correlation functions in the continuum with functional methods. To that end, in this work we employ the fRG which has been extensively developed in this context. 
Examples include studies of colour confinement~\cite{Ellwanger:1995qf,Gies:2002af, Pawlowski:2003hq,Cyrol:2016tym,Cyrol:2017qkl,Corell:2018yil,Ferreira:2025tzo}, dynamical chiral symmetry breaking~\cite{Gies:2001nw, Gies:2002hq,Braun:2011pp,Mitter:2014wpa,Cyrol:2017ewj}, full QCD at finite $T$ and $\mu_{\rm B}$~\cite{Fu:2019hdw,Ihssen:2024miv}, and the many-flavour regime~\cite{Gies:2005as,Braun:2006jd,Braun:2009ns,Braun:2010qs,Goertz:2024dnz}.

In the following sections, we present results for the thermal chiral effective potential computed with the fRG, using quantitative approximations that have been tested in the context of physical QCD~\cite{Ihssen:2024miv}. 
In this QCD-like framework, the scalar-pseudoscalar $(\sigma\!-\!\pi)$ channel is bosonised, capturing the dynamics of the lightest composite degrees of freedom. Together with the fermions, these drive the deep IR dynamics relevant for the phase structure and precondensation. 
Moreover, by bosonising only this channel we impose maximal U(1)$_{\rm A}$-breaking, for which the $(\eta\!-\!a)$ tensor structures are relatively suppressed. This is a good approximation in the few-flavour limit. In turn, the axially-symmetric approximation, as used in \cite{Goertz:2024dnz}, is better suited for larger $N_f$: the momentum dimension of the ’t~Hooft operator increases with $N_f$ and hence its dynamical relevance decreases. 
For more details on the implementation of gauge-fermion theories using the fRG, we refer to \Cref{app:Computation} and~\cite{Ihssen:2024miv, Goertz:2024dnz}.

%%%%%%%%%%%%%%%%%%%%%%%%%%%%%%%%%%%%%%%%
\subsection{$\dSSB$ and precondensation}
\label{sec:PrecondGaugeFermion}
\begin{figure}[t!]
	\centering
	\includegraphics[width=.925\columnwidth]{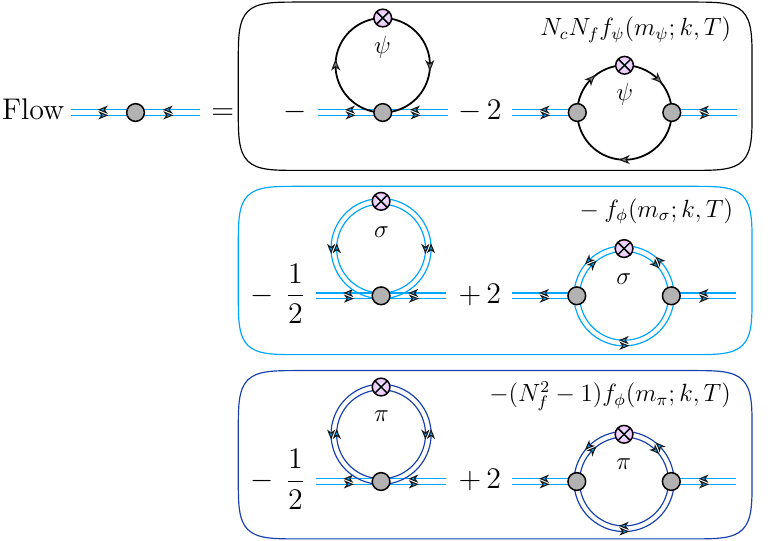}
	\caption{Diagrammatic flow of the radial $\sigma$-mode two-point function, which encodes the curvature of the effective potential at small fields and thus the emergence of a condensate. Blue, double-arrowed lines denote full propagators of the $\sigma$ and $\boldsymbol{\pi}$ modes (dark blue), while single-arrowed lines represent fermion propagators. Gray dots indicate full vertices, and crossed circles denote insertions of the scale derivative of the regulator function, as usual in the fRG.}
	\label{fig:Flow_Gsigma}
\end{figure}

The dynamics responsible for precondensation is the same as the one driving $\dSSB$, where a fermionic condensate forms and bound states emerge. 
From a fundamental perspective, the relevant information is encoded in fermionic self-interactions, which can be mapped by a field redefinition onto a Yukawa-type theory that provides an efficient low-energy description. 
Within the framework of the fRG, this is successfully implemented through a scale-dependent field redefinition~\cite{Gies:2001nw,Gies:2002hq,Pawlowski:2005xe} which allows to continuously account for quantum fluctuations from the fundamental UV down to the deep IR, beyond the $\dSSB$ scale where the dynamics are governed by the emergent composites.  
This setup also permits a systematic inclusion of fermionic self-interactions of arbitrarily high order in the bosonised channel, capturing genuinely non-perturbative effects and leading to a robust truncation of the fermion sector.  
In this formulation, all non-trivial IR dynamics are uniquely determined by the UV properties of the theory, which emphasises the first-principles character of the approach.

To describe $\dSSB$, it suffices to discuss the quantum corrections in the flow of the mesonic effective potential $V_\textrm{eff}(\phi^2)$ for the average bosonised field $\phi = (\sigma,\,\boldsymbol{\pi})$, where the radial $\sigma$-mode is the direction of chiral symmetry breaking and $\boldsymbol{\pi}$  are $N_f^2-1$ pseudoscalar exact Goldstone bosons.
SSB then occurs when quantum corrections induce a new vacuum breaking a classical symmetry of the theory. In the present case, a condensate $\sigma_0$ develops in the radial direction, providing a mass to the fermions and dynamically breaking the chiral symmetry.

The phase transition can be studied along the RG-flow through the scale dependence of $V_\textrm{eff}(\phi^2)$ and, particularly for the case of a second order phase transition, the emergence of a new vacuum is given by the sign change of curvature around the symmetric trivial minimum at $\sigma_0=0$ which is given by the $\sigma$-mode mass, $m_\sigma^2(\phi^2) = \partial_\sigma^2 V_{\rm eff}(\phi^2)$.
With this, the scale dependence of the condensate $\sigma_0(k)$ is obtained and can be related to $\sigma_0(x;r=k^{-1})$ as introduced in \Cref{sec:PrecondensationBasics}.

The relevant quantum corrections are then those to $m_\sigma^2$ and encoded in the flow of the corresponding inverse propagator $G_\sigma^{-1}(p)$,
\begin{align}
    k\partial_k \,G_\sigma^{-1}(p) &= \adjincludegraphics[height=0.85em,valign=t]{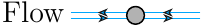}
    \notag\\[2.ex]
    &= N_cN_ff_\psi(m_\psi,p;k,T) 
    \notag\\[0.5ex]
    &\qquad- f_\phi(m_\sigma,p;k,T) 
    \notag\\[0.5ex]
    &\qquad- (N_f^2-1)f_\phi(m_\pi,p;k,T)\,.
    \label{eq:flowInvProp}
\end{align}
Here, we have introduced the functions $f_{\psi/\phi}$, which contain non-perturbative and thermal corrections in different forms (see \Cref{app:GsigmaFlow} for the explicit expressions), as a shorthand notation for the classes of diagrams depicted in each of the boxes in~\Cref{fig:Flow_Gsigma}. 
At $\sigma=0$, all $f_i > 0$ in the relevant regime, as we show in \Cref{app:GsigmaFlow}.
In each of these functions, two diagrammatic topologies contribute: the tadpole (with one four-point function) and the polarisation diagram (with two three-point functions), each mediated by fermion, $\sigma$, or $\boldsymbol{\pi}$ loops.
The factors of $N_f$ and $N_c$ follow from the multiplicity of the respective loops, which depends on the type of propagating off-shell particle.

The overall sign of the fermionic contribution is positive due to their Grassmannian nature, which introduces a natural competition among fermionic and (negative) bosonic corrections. 
To be explicit, the positive sign of the fermionic contribution drives the curvature negative, enhancing $\dSSB$, whereas the bosonic contribution opposes this.

At very high temperatures, fermionic fluctuations fully dominate over the bosonic ones. Still, their contributions are strongly suppressed by their large thermal mass, ${m_\psi \sim \pi T}$, and are therefore insufficient to induce a negative curvature. 
The theory thus remains in the symmetric regime. As the temperature decreases to $T \sim \Tpre$, the thermal mass of the fermion becomes small enough to allow for a sign change in $m_\sigma^2$, and the emergence of a non-trivial condensate. This occurs at the momentum scale $k = r_{\rm UV}^{-1}$, indicated by the blue line in \Cref{fig:OrderParameter}.

At low temperatures, where $\dSSB$ occurs, the pions turn massless, being exact Nambu-Goldstone modes, and counteract the fermionic contributions. Once a condensate has formed, this effect is mainly driven by the purely bosonic polarisation diagrams shown in the second row of \Cref{fig:Flow_Gsigma}, which dominate due to their large multiplicity $ N_f^2-1$.
Although fermionic corrections are suppressed at very small $k$ due to the finite mass generated by the scalar condensate, the effect of the bosons is still weaker and they cannot restore the symmetry.

At intermediate temperatures, the difference in finite-temperature effects on fermions and bosons becomes relevant: while  the fermions simply obtain a thermal mass, the bosons retain a zero mode (with no direct thermal mass), while higher frequencies decouple. This is the phenomenon of dimensional reduction. 
Thus, for $T \lesssim \Tpre$ bosonic fluctuations are still enhanced at $k \lesssim r_{\rm UV}^{-1}$ (blue line in \Cref{fig:OrderParameter}) and overtake the gapped fermions to restore the symmetry at $k = \xi^{-1}$ (red line in \Cref{fig:OrderParameter}). This leads to the presence of a condensate only at a finite range of length-scales, which is the hall-mark of precondensation.

With this discussion we illuminated the microscopic origin of precondensation as a competition of fermionic and bosonic fluctuations, as well as thermal effects. In summary, the underlying mechanism requires some key ingredients:
\begin{enumerate}
	\item[\textbf{(i)}] \textbf{{Competing counteracting effects}},
	which enhance and suppress $\dSSB$ are indispensable for the restoration of the symmetry at finite length scales. These occur naturally in fermionic theories with chiral symmetry breaking.
    Though we have discussed this at the example of the QCD-like theories, analogous mechanisms are present in two-dimensional Dirac semimetals described by the Gross–Neveu model~\cite{Tolosa-Simeon:2025fot}, or in cold-atom systems near the BCS–BEC crossover, where the interplay between fermions and Cooper pairs plays a similar role~\cite{Boettcher:2012cm,Boettcher:2012dh,Boettcher:2013kia}.
    Interestingly, a similar behaviour can be also engineered in purely scalar theories through portal couplings~\cite{Hawashin:2024dpp,Goertz:2023pvn}. 
	
	\item[\textbf{(ii)}] \textbf{{Finite-temperature effects}} or possibly other external physical parameters (e.g.,~the chemical potential) which introduce an additional scale and change the relevance of the two competing species. The resulting relative thermal suppression of fermionic and enhancement of bosonic corrections is crucial for precondensation. For instance, the same system as here was studied in the vacuum in~\cite{Goertz:2024dnz}, where no explicit signatures of precondensation could be found, given the absence of thermal fluctuations. 

	\item[\textbf{(iii)}] \textbf{Massless modes} are necessary to restore the symmetry at momentum scales $k \lesssim r_{\rm UV}^{-1}$. In the presence of a finite mass for the pionic modes (e.g. by explicit symmetry breaking of the classical symmetry), the fluctuations responsible for the restoration decouple at scales $k \sim m_\pi$. Consequently, precondensation can be viewed as a relatively generic feature of phase transitions with exact Goldstone bosons and is therefore expected in other systems with chiral symmetry, see e.g.~\cite{Li:2025tvu,Janssen:2014gea}.
    An interesting open question concerns the fate of precondensation features, such as condensate modulations, in the presence of a small explicit symmetry breaking. 
    In this case, one expects the modulation of the condensate to persist, in close analogy with Weiss domains in the presence of a weak global magnetisation, and to be progressively washed out as the pion mass $m_\pi$ is increased.
\end{enumerate}
%

%%%%%%%%%%%%%%%%%%%%%%%%%%%%%%%%%%%%%%%%
\subsection{Many-flavour scaling}
\label{sec:Nfscaling}

\begin{figure}[t!]
	\centering
	\includegraphics[width=.95\columnwidth]{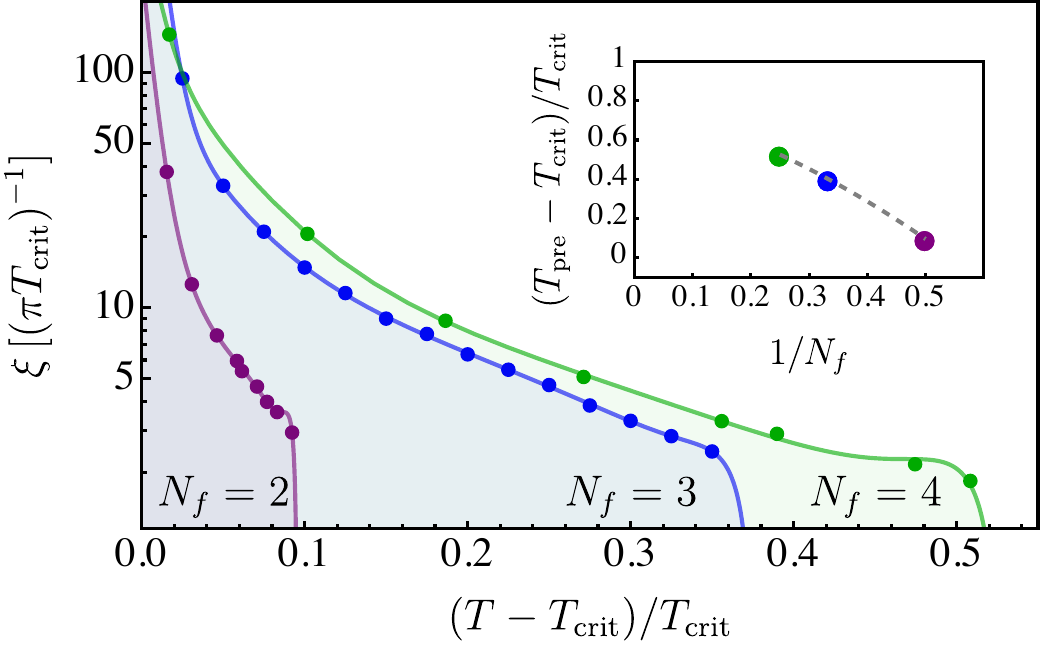}
	\caption{
    Domain size as a function of the temperature relative to the critical one. As purple, blue and green dots the results for $N_f=2$, $3$ and $4$ are respectively shown which can also be found in \Cref{fig:OrderParameter,fig:precondensationNf2andNf4}. In the inlay plot we show the reduced precondensation temperature as a function of $N_f^{-1}$. 
    }
	\label{fig:domainsizeNfscaling}
\end{figure}

An important question in gauge-fermion theories in the chiral limit, and in particular in those with QCD-like symmetries, concerns the thermal properties as $N_f$ is increased. 
This issue is relevant for the theoretical understanding of their phase structure and, from a phenomenological perspective, for assessing their viability and detectability as extensions of the Standard Model of particle physics.

\begin{figure*}[t!]
	\centering
	\includegraphics[width=2\columnwidth]{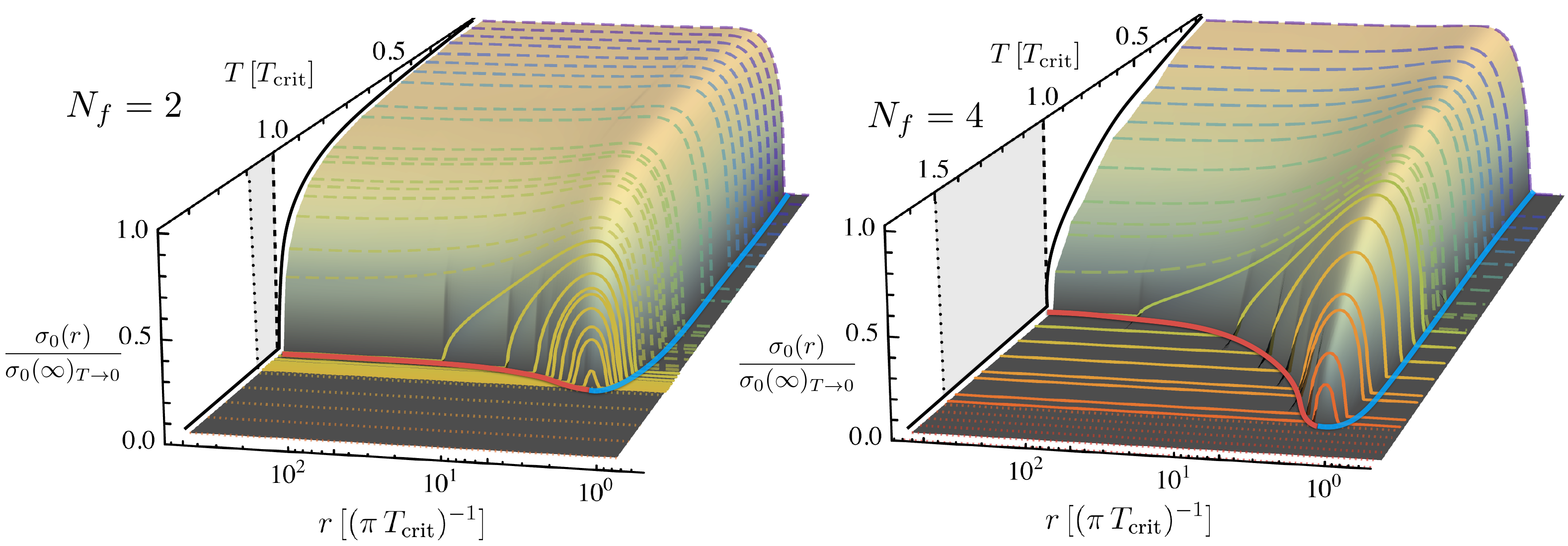}
	\caption{Chiral order parameter $\sigma_0(r)$ of two QCD-like gauge-fermion theories at $N_c=3$, with $N_f=2$ (left) and  $N_f=4$ (right) chiral flavours as a function of the spatial separation $r=k^{-1}$ and temperature, normalised to its value at zero temperature in the macroscopic limit ($\sigma_0(\infty)_{\,T\to0}$). Labels are the same as in \Cref{fig:OrderParameter}.
    }
	\label{fig:precondensationNf2andNf4}
\end{figure*}

We have studied theories with different flavour numbers $N_f$ while keeping $N_c=3$ fixed. We show the resulting chiral condensates for $N_f=2,\, 3$ and $4$ in \Cref{fig:OrderParameter,fig:precondensationNf2andNf4}. 
As the pion multiplicity grows quadratically with $N_f$ (see \labelcref{eq:flowInvProp}), the restoration of chiral symmetry in the precondensation phase is enhanced as $N_f$ increases. This delays the onset of a macroscopic condensate and extends the precondensation regime over a broader range of temperatures and scales. We illustrate this in \Cref{fig:domainsizeNfscaling}, where the size of the precondensation domains is shown as a function of temperature in units of the respective $T_c$. As expected, the precondensation dynamics become more relevant for larger $N_f$. 

The order of the thermal chiral phase transition in QCD-like theories in the chiral limit is still an intensely studied open question. For small $N_f$ the transition is of second order, and it may change into a first order one with growing $N_f$; see e.g.~\cite{Resch:2017vjs, Klinger:2025xxb, Cuteri:2021ikv} for discussions in this context. 
The axial anomaly plays a crucial rôle in the respective dynamics, as it adjusts the relevance of the scalar and pseudo-scalar $(\eta\!-\!a)$ modes. Hence, it is expected to drive the change in the order of the transition, see e.g.~\cite{Resch:2017vjs, Fejos:2014qga, Fejos:2018dyy, Fejos:2021yod, Fejos:2022mso, Fejos:2024bgl}. The present framework considers maximal breaking of axial symmetry, which is an optimal approximation in the few-flavour regime. While the results obtained here display a second-order transition, in the many-flavour regime, the momentum dimension of the ’t~Hooft operator increases with $N_f$, leading to a respective degeneracy between the scalar $\sigma$ and $a$, as well as the pseudoscalar $\pi$ and $\eta$ channels. This would effectively double the number of dynamical modes, which could drive the change in the order of the chiral phase transition. 

The impact of the axial $(\eta\!-\!a)$ modes on the dynamics of precondensation can be estimated. 
These modes contribute, in addition to the axially symmetric ones shown in \Cref{fig:Flow_Gsigma}, with $N_f^2-1$ additional scalar channels ($a$) and one pseudoscalar ($\eta$). 
While the former contribute as the $\sigma$ mode and are expected to decouple from the low-energy dynamics, the latter add to the Goldstone modes. 
Consequently, given the same sign of these corrections and their multiplicity, it is reasonable to expect that precondensation is mildly enhanced in the many-flavour limit by the restoration of the axial~symmetry.

%%%%%%%%%%%%%%%%%%%%%%%%%%%%%%%%%%%%%%%%
\section{Conclusions}
\label{sec:Conclusions}

We have studied precondensation in thermal gauge-fermion theories in the chiral limit for $N_c = 3$ and $N_f = 2,\,3,\,4$. This phenomenon arises from the interplay between fermionic fluctuations and those of bosonic bilinear composites. 
While the former drive chiral symmetry breaking, the latter are symmetry-restoring. Their relative importance changes at finite temperature in a momentum-dependent manner, which is key to the emergence of momentum-dependent condensates. 

We infer from our results that the temperature window in which precondensation occurs increases rapidly with $N_f$, see \Cref{fig:OrderParameter}, \ref{fig:domainsizeNfscaling} and \ref{fig:precondensationNf2andNf4}. 
This suggests that precondensation becomes increasingly relevant in the many-flavour regime and may significantly affect the near-critical dynamics of the chiral phase transition. In this work, we have assumed maximal axial symmetry breaking induced by topological effects, which is a reliable approximation at small~$N_f$. For~$N_f\gtrsim 4$, the inclusion of the $(\eta$--$a)$ modes is of increasing importance for the fate of precondensation and the order of the thermal transition. Importantly, these modes only add to the competing dynamics and hence, precondensation is expected to be a generic feature of chiral theories with a phase transition. 

The mechanism underlying precondensation is not specific to gauge-fermion systems, but also occurs in general condensed-matter and cold-atom settings. 
Precondensation requires the presence of two key ingredients:
\begin{itemize}
\item[(1)] competing symmetry-breaking and symmetry-restoring dynamics from different degrees of freedom at the level of quantum corrections and, 
\item[(2)] an external tuning parameter, such as temperature or density, that can be used to reweight the relative importance of degrees of freedom in the momentum-dependent way. 
\end{itemize}
It follows from these two very general dynamical foundations that precondensation is not restricted to thermal phase transitions. It can arise under variations of other external parameters, such as density or system size.
Additionally, (1) and (2) also entail that precondensation is closely related to many other interesting phenomena such as pseudo-gapped phases, the emergence of moat regimes or even inhomogeneous condensates, as these phenomena also exhibit a non-trivial momentum dependence of the condensate.

Gauge-fermion theories of the type discussed here are  candidates for physics beyond the Standard Model. 
They may help to address  long-standing open questions such as the hierarchy problem and the nature of dark matter, and precondensation, as well as the related phenomena mentioned above may play a key rôle. We hope to report on these topics in the near future.

%%%%%%%%%%%%%%%%%%%%%%%%%%%%%%%%%%%%%%%%
\subsection*{Acknowledgements}

We thank Mark G. Alford, Adrien Florio, Friederike Ihssen, Michael Scherer, Mireia Tolosa-Sime\'on, Shahram Vatani, Jonas Wessely and Masatoshi Yamada for discussions.
This work is done within the fQCD collaboration \cite{fQCD} and we thank its members for discussions and collaborations on related projects.
APG is supported by the RIKEN Special Postdoctoral Researcher (SPDR) Program.
FRS acknowledges funding by the GSI Helmholtzzentrum f\"ur Schwerionenforschung and by HGS-HIRe for FAIR.
FRS is supported by the Deutsche Forschungsgemeinschaft (DFG, German Research Foundation) through the Emmy Noether Programme Project No. 54526179.
FRS also thanks RIKEN iTHEMS, for the kind hospitality and support during his stay.

%%%%%%%%%%%%%%%%%%%%%%%%%%%%%%%%%%%%%%%%%%
\section*{Appendix}
\appendix
\begingroup
\allowdisplaybreaks

\crefalias{section}{appendix}
\crefalias{subsection}{appendix}
%%%%%%%%%%%%%%%%%%%%%%%%%%%%%%%%%%%%%%%%%

%%%%%%%%%%%%%%%%%%%%%%%%%%%%%%%%%%%%%%%%
\section{Computational details}
\label{app:Computation}

In this Appendix, we provide some details on the theoretical setup and computational methodology. The presently employed framework has been developed in previous works and tested for quantitative accuracy in physical QCD, see~\cite{Sattler:2025hcg,Ihssen:2024miv,Pawlowski:2025jpg}. Here, we provide only the most relevant aspects thereof. 

In our explicit computation, the Landau gauge limit is taken, i.e. $\xi\to 0$, and we restrict ourselves to $N_c=3$ and $N_f=2,3,4$ massless fermions. These theories are close cousins of physical QCD and hence we refer for the systematic error analysis to \cite{Ihssen:2024miv} and references therein. For $N_f< 11/2\, N_c$, see \cite{Goertz:2024dnz}, the theory is asymptotically free and features confinement and $\dSSB$ in the IR. Moreover, for few flavours $N_f=2,3,4$, the respective confinement and chiral scales are close to each other \cite{Goertz:2024dnz}.

%%%%%%%%%%%%%%%%%%%%%%%%%%%%%%%%%%%%%%%%
\subsection{Numerical implementation} 
\label{app:numerics}

We compute our results from a first-principles approach to QCD, using an fRG setup corresponding to the one used in \cite{Ihssen:2024miv}, but improved to include full momentum dependences in the meson and fermion sectors, and expanded to finite temperatures. This setup has also been employed in \cite{Pawlowski:2025jpg}. An in-detail description thereof may be found in Chapter~9 and Appendix~H of~\cite{Sattler:2025hcg}.
Here, we just briefly sketch the setup:\\[-2ex]
\begin{itemize}
	\item We calculate the full glue sector with an identification of the RG-scale and the physical scale~${k=p}$. For the pure glue propagators, we calculate temperature-dependent corrections for both glue and ghost propagators on top of vacuum data from \cite{Cyrol:2017ewj}. All other correlation functions are computed self-consistently at each temperature. In particular, we also explicitly parametrise the gluon propagator with a flowing gluon mass gap, which carries most of the thermal effects on the gluon propagator.
	\item The Polyakov-loop potential $V_\textrm{glue}(A_0)$ is evaluated similarly to \cite{Fu:2019hdw}: we take pure Yang-Mills input from \cite{Lo:2013hla} and add to it the quark-corrections from our flows.
	\item In the low-energy sector, consisting of quarks and mesons, we calculate fully momentum-dependent propagators for the mesons and quarks. To compute the fully field-dependent effective potential $V_\textrm{eff}(\phi^2)$, describing multi-meson scatterings, we utilise methods from numerical hydrodynamics, see \cite{Grossi:2019urj, Grossi:2021ksl, Ihssen:2022xkr, Ihssen:2023qaq, Ihssen:2023xlp, Ihssen:2024miv}.
\end{itemize}
All calculations are performed using a 3D regulator with the shape function $s_8$, as defined in Chapter~3.4.3 of~\cite{Sattler:2025hcg}.
Our flow equations have been derived with the help of \texttt{QMeS}\cite{Pawlowski:2021tkk} and the resulting diagrams are traced using \texttt{FormTracer}\cite{Cyrol:2016zqb}. Tensorial bases and the projection operators for all vertices are derived using \texttt{TensorBases}\cite{Braun:2025gvq} and the resulting flow equations are solved in the high-performance C++ framework of \texttt{DiFfRG}\cite{Sattler:2024ozv}.

%%%%%%%%%%%%%%%%%%%%%%%%%%%%%%%%%%%%%%%%
\subsection{Flow initialisation}
\label{app:InitialCondition} 

We initialise all of our RG-flows at an initial scale of $\Lambda_\textrm{UV}~=~20$~GeV and flow down to a scale $k \approx 10^{-4}\,\textrm{GeV}$, where all chiral observables have settled. 
As noted in \cite{Ihssen:2024miv}, in the current setup one requires a slight "de-hancement" $a$ for the quark-gluon vertex to obtain a quark mass of $370$ MeV at a pion mass of $138\textrm{ MeV}$. 
In practice, we slightly decrease the flow of $\alpha_{q\bar{q}A}$ by $a=4\%$ below $k=2\textrm{ GeV}$, also at $m_\pi = 0$, where our simulations are performed at.

For the other $N_f=3,4$, we take exactly the same "de-hancement" factor, but adjust the initial values of the avatars of the strong interaction such that we fulfil the perturbative STIs, i.e.
\begin{align}
	\alpha_{c\bar{c}A}(p) = \alpha_{A^3}(p) = \alpha_{A^4}(p) = \alpha_{q\bar{q}A}(p)\,,
\end{align}
at $p\gtrsim5\textrm{ GeV}$. As argued in \cite{Ihssen:2024miv}, the RG-scale $k$ is a good proxy for the physical momentum $p$, and we fix the STIs at the level of $k$ instead of $p$.\\
Note that except for the "de-hancement" $a$, no further parameters enter our calculation. In particular, the full meson sector naturally emerges from the quark-gluon dynamics at high momenta via our implementation of emergent composites, see \cite{Gies:2001nw, Pawlowski:2005xe, Ihssen:2024miv}.

%%%%%%%%%%%%%%%%%%%%%%%%%%%%%%%%%%%%%%%%
\section{Flow of $G_\sigma^{-1}$} 
\label{app:GsigmaFlow}

In the following, we list the diagrammatic parts of the flow equation for the inverse scalar meson propagator $G_\sigma^{-1}(p)$ as introduced in \Cref{sec:PrecondGaugeFermion}, see also \Cref{fig:Flow_Gsigma}.
To shorten and simplify the presentation, we only provide the functions $f_i$ at $\sigma = 0$ and~$p=0$,
\begin{widetext}
\begin{align}
    N_cN_f f_\psi &= N_c N_f\sum_{q_0\in\pi T(2\mathbb{Z}+1)}\int \frac{d^3 q}{(2\pi)^3}\,\,2\,
    h_{\sigma \psi  \bar{\psi }}\left(0,q\right) G_{\psi }\left(q\right) h_{\sigma\psi\bar{\psi}}\left(0,q\right)G_{\psi }\left(q\right)\dot{R}_F\left(q\right) G_{\psi }\left(q\right)
    \notag\\&\hspace{19em}\times
\left(|\boldsymbol{q}| + R_F\left(q\right)\right)
\left(\left(R_F\left(q\right)+|\boldsymbol{q}|\right)^2+q_{0}^2\right)\,,
\notag\\[1.5ex]
    -(N_f^2-1) f_\pi &= -\frac{1}{2} (N_f^2-1) 
    \sum_{q_0\in2\pi T\mathbb{Z}}\int \frac{d^3 q}{(2\pi)^3}\,
    \partial_\rho^2V_{\textrm{eff}}(0)
    \,G_\pi\left(q\right) \dot{R}_B(q) G_\pi\left(q\right)\,,
\notag\\[1.5ex]
    -f_\sigma &= -\frac32
    \sum_{q_0\in2\pi T\mathbb{Z}}\int \frac{d^3 q}{(2\pi)^3}\,
    \partial_\rho^2V_{\textrm{eff}}(0)
    \,G_\sigma(q)\dot{R}_B(q) G_\sigma(q)\,.
    \label{eq:GsigmaFlows}
\end{align}
\end{widetext}
In the above, we have also used the $O(N_f^2-1)$ invariant $\rho = \frac{\sigma^2+\boldsymbol{\pi}}{2}$ and $q = (q_0, \boldsymbol{q})$. Furthermore, in the above formula, for brevity we have set the fermionic wave-functions $Z_\psi^\parallel \approx Z^\perp_\psi\approx1$, which does not impact the qualitative features of \labelcref{eq:GsigmaFlows}.

As argued before, to diagnose whether $\dSSB$ occurs we only need to follow the flow of $G^{-1}(0)$ at $\sigma=0$ and monitor it for a change of sign.
In the above, we have used $\rho = {(\sigma^2+\boldsymbol{\pi}^2)}/{2}$, where $h_{\sigma\psi\bar\psi}$ is the Yukawa coupling between the scalar mode and the quarks, and $G_{\pi/\sigma/\psi}$ are the scalar parts of the respective~propagators,
\begin{align}
    G_i(p) &= \frac{(Z_i^\parallel)^{-1}}{p_0^2 + z_i^\perp(p)\boldsymbol{p}^2 + m_i^2 + R_B(\boldsymbol{p})}
    \,,\qquad\text{for } i=\sigma,\pi\,,
    \notag\\[1ex]
    G_\psi(p) &= \frac{1}{Z_\psi^\parallel\,p_0^2 + m_\psi^2 + Z_\psi^\perp(p)(|\boldsymbol{p}| + R_F(\boldsymbol{p}))}\,.
\end{align}
Note that the fermionic tadpole, as well as the purely bosonic polarisation diagrams are identical to $0$ at $\rho = 0$ and therefore drop out in \labelcref{eq:GsigmaFlows}. With this discussion, it is apparent that
\begin{align}
    f_\psi > 0\,,\qquad
    f_\pi > 0 \qquad {\rm and }\quad
    f_\sigma > 0\,,
\end{align}
which we have used in \Cref{sec:GaugeFermion} to explain the competition between fermionic and bosonic degrees of freedom.

%%%%%%%%%%%%%%%%%%%%%%%%%%%%%%%%%%%%%%%%
\section{More details on local order and the renormalisation group} 
\label{app:fRG}

In this Appendix, we discuss the relation between the spatial average in $\sigma^2(\boldsymbol{x};r)$ and the cutoff-dependent condensate $\sigma_k$. 
For this purpose, we start by inserting a regulator function in position space for the cutoff term in \labelcref{eq:RegPImeasure}, instead of using one in momentum space. 
Moreover, for the sake of simplicity, we restrict ourselves to a sharp distance cutoff $R_r(\boldsymbol{z})$ with $\boldsymbol{z}= \boldsymbol{x}-\boldsymbol{y}$, where $R_r(\boldsymbol{z})$ vanishes for $|\boldsymbol{z}|<r$ and diverges for $|\boldsymbol{z}|>r$. 
With this cutoff, we remove all modes from the path integral that are correlated over a distance larger than $r$. 
In the presence of such a cutoff term, \labelcref{eq:Condensate-r} reduces to 
\begin{align} \nonumber 
	\sigma_0^2(\boldsymbol{x};r) = &\,\frac{1}{{\cal V}_r}\int d^3 y \,\langle \hat\sigma(\boldsymbol{x})\,\chi_r(\boldsymbol{z})\,\hat\sigma(\boldsymbol{y})\rangle \\[1ex] 
    = &\, \frac{1}{{\cal V}_r}\int d^3 y \,\langle \hat\sigma(\boldsymbol{x})\,\hat\sigma(\boldsymbol{y})\rangle\,,
	\label{eq:RegSpatial}
\end{align} 
where $\chi_r(\boldsymbol{z})=\theta(r^2-\boldsymbol{z}^2)$ is the  characteristic function of all modes which are \textit{not} removed by the sharp regulator function $R_r(\boldsymbol{z})$. 
In short, the sharp spatial distance cutoff term implements precisely the averaging procedure put forward in \Cref{sec:PrecondensationBasics}.
Note that the average $\langle\cdot\rangle$ in the above is already performed with respect to the regulated path integral.

In the presence of domains or further spatial modulations, \labelcref{eq:RegSpatial} may have a dependence on the position $\boldsymbol{x}$. 
This dependence is removed by a further average over $\boldsymbol{x}$, to wit, 
\begin{align} 
	\sigma_0^2(r) = \frac{1}{{\cal V}\,{\cal V}_r}\int d^3x\,d^3 y \,\langle \hat\sigma(\boldsymbol{x})\,\chi_r(\boldsymbol{z})\,\hat\sigma(\boldsymbol{y})\rangle \,. 
\label{eq:RegSpatialFullAv}
\end{align} 
The final step concerns the translation of this relation to the fRG with momentum-space cutoffs as used in our explicit computation, for details see Chapter~3.4.3 of~\cite{Sattler:2025hcg}. 
To that end we rewrite \labelcref{eq:RegSpatialFullAv} in momentum space which leads us to 
\begin{align} 
\sigma_0^2(r) =  \frac{1}{{\cal V}}\int \frac{d^3 q}{(2\pi)^3}\,\langle \hat\sigma(-\boldsymbol{q})\,\tilde \chi_r(\boldsymbol{q})\,\hat\sigma(\boldsymbol{q})\rangle \,, 
\label{eq:RegSpatialFullAvp}
\end{align} 
with 
\begin{align} 
\tilde \chi_r(\boldsymbol{q} )= \frac{1}{{\cal V}_r} \int d^3 z\,  \chi_r(\boldsymbol{z})\,e^{\imag \boldsymbol{z} \boldsymbol{q}}\,. 
\label{eq:tildechi}
\end{align}
The normalised Fourier transform of the characteristic function $\chi_r(\boldsymbol{z})$ is a smeared out characteristic function in momentum space: we have ${\tilde \chi_r(0)=1}$ and ${\tilde \chi_r(\boldsymbol{q}^2\to \infty)=0}$, with the transition happening at ${\boldsymbol{q}^2\propto 1/r^2}$. 

For a closer analogy to the regulator functions we employed in this work, we now compare \labelcref{eq:RegSpatialFullAvp} with a similar expectation value in the presence of a sharp IR momentum cutoff obtained from \labelcref{eq:RegPImeasure}, with the regulator function 
\begin{align} 
R_k^{\textrm{(sharp)}}(\boldsymbol{p}) = k^2 \left( \frac{1}{\theta(\boldsymbol{p}^2-k^2} -1\right)\,.
\label{eq:Rsharp}
\end{align}
Then we are led to 
\begin{align}
    \frac{1}{{\cal V}}\int  d^3 q \,\langle \hat\sigma(- {\bm q})\,\hat\sigma({\bm q}) \rangle = \frac{1}{{\cal V}}
    \int  d^3 q \,\langle \hat\sigma( - {\bm q})\,\chi_k(\boldsymbol{q})\hat\sigma({\bm q})\rangle \,,
    \label{eq:2pointMomentumCutoff}
\end{align}
with the characteristic momentum mode function ${\chi_k(\boldsymbol{p})=\theta(k^2-\boldsymbol{p}^2)}$. This is simply \labelcref{eq:RegSpatialFullAvp} with ${\tilde \chi_r(\boldsymbol{q})\to \chi_k(\boldsymbol{q})}$: we have substituted the smooth version $\tilde \chi_r(\boldsymbol{q})$ of the characteristic function with a sharp one in momentum space with $k\sim 1/r$. 

Evidently, a suitably weighted Fourier transform of the right-hand side of \labelcref{eq:RegSpatialFullAvp} to position space results in \labelcref{eq:RegSpatialFullAv}, with ${\chi_r(\boldsymbol{z})\to \tilde \chi_k(\boldsymbol{z})}$, 
\begin{align} 
    \tilde \chi^{\ }_k(\boldsymbol{z} )= \frac{1}{{\cal V}_{k}}\int \frac{d^3 q}{(2\pi)^3}\,  \chi^{\ }_k(\boldsymbol{q})\,e^{-\imag \boldsymbol{z} \boldsymbol{q}}\,, 
\label{eq:tildechik}
\end{align}
where ${\cal V}_k =1/ {\cal V}_{r=1/k}$ is proportional to the spatial momentum volume of $\boldsymbol{q}^2\leq k^2$. This normalisation is chosen such that $\tilde{\tilde\chi}_r(z)=\chi_r(z)$.

In summary, this allows us to define the analogue of \labelcref{eq:Condensate-rp} with a sharp momentum cutoff is obtained by simply substituting $\delta_r(\boldsymbol{q})\to \tilde \chi_k(\boldsymbol{q})$ and ${\cal V}_r \to {\cal V}_{1/k}$, 
\begin{align} 
	\sigma_0^2({\bm p};k^{-1}) = \frac{1}{{\cal V}_{k^{-1}}}\int  d^3 q \,\langle\hat\sigma({\bm p} - {\bm q})\,\tilde \chi^{\ }_k(\boldsymbol{q})\,\hat\sigma({\bm q})\rangle\,. 
	\label{eq:CondensateRegMomentum}
\end{align}
Note that $\sigma_0^2(r) = \sigma_0^2(0;k^{-1})/{\cal V}$, and hence \labelcref{eq:CondensateRegMomentum} underlies our analysis of the length-dependent condensate in \Cref{sec:GaugeFermion}.

%%%%%%%%%%%%%%%%%%%%%%%%%%%%%%%%%%%%%%%%
\bibliography{references}

\begin{thebibliography}{82}%
\makeatletter
\providecommand \@ifxundefined [1]{%
 \@ifx{#1\undefined}
}%
\providecommand \@ifnum [1]{%
 \ifnum #1\expandafter \@firstoftwo
 \else \expandafter \@secondoftwo
 \fi
}%
\providecommand \@ifx [1]{%
 \ifx #1\expandafter \@firstoftwo
 \else \expandafter \@secondoftwo
 \fi
}%
\providecommand \natexlab [1]{#1}%
\providecommand \enquote  [1]{``#1''}%
\providecommand \bibnamefont  [1]{#1}%
\providecommand \bibfnamefont [1]{#1}%
\providecommand \citenamefont [1]{#1}%
\providecommand \href@noop [0]{\@secondoftwo}%
\providecommand \href [0]{\begingroup \@sanitize@url \@href}%
\providecommand \@href[1]{\@@startlink{#1}\@@href}%
\providecommand \@@href[1]{\endgroup#1\@@endlink}%
\providecommand \@sanitize@url [0]{\catcode `\\12\catcode `\$12\catcode
  `\&12\catcode `\#12\catcode `\^12\catcode `\_12\catcode `\%12\relax}%
\providecommand \@@startlink[1]{}%
\providecommand \@@endlink[0]{}%
\providecommand \url  [0]{\begingroup\@sanitize@url \@url }%
\providecommand \@url [1]{\endgroup\@href {#1}{\urlprefix }}%
\providecommand \urlprefix  [0]{URL }%
\providecommand \Eprint [0]{\href }%
\providecommand \doibase [0]{http://dx.doi.org/}%
\providecommand \selectlanguage [0]{\@gobble}%
\providecommand \bibinfo  [0]{\@secondoftwo}%
\providecommand \bibfield  [0]{\@secondoftwo}%
\providecommand \translation [1]{[#1]}%
\providecommand \BibitemOpen [0]{}%
\providecommand \bibitemStop [0]{}%
\providecommand \bibitemNoStop [0]{.\EOS\space}%
\providecommand \EOS [0]{\spacefactor3000\relax}%
\providecommand \BibitemShut  [1]{\csname bibitem#1\endcsname}%
\let\auto@bib@innerbib\@empty
%</preamble>
\bibitem [{\citenamefont {Tolosa-Sime{\'o}n}\ \emph {et~al.}(2025)\citenamefont
  {Tolosa-Sime{\'o}n}, \citenamefont {Classen},\ and\ \citenamefont
  {Scherer}}]{Tolosa-Simeon:2025fot}%
  \BibitemOpen
  \bibfield  {author} {\bibinfo {author} {\bibfnamefont {M.}~\bibnamefont
  {Tolosa-Sime{\'o}n}}, \bibinfo {author} {\bibfnamefont {L.}~\bibnamefont
  {Classen}}, \ and\ \bibinfo {author} {\bibfnamefont {M.~M.}\ \bibnamefont
  {Scherer}},\ }\href {\doibase 10.1103/7kw4-8r3m} {\bibfield  {journal}
  {\bibinfo  {journal} {Phys. Rev. B}\ }\textbf {\bibinfo {volume} {112}},\
  \bibinfo {pages} {115133} (\bibinfo {year} {2025})},\ \Eprint
  {http://arxiv.org/abs/2503.04911} {arXiv:2503.04911 [cond-mat.str-el]}
  \BibitemShut {NoStop}%
\bibitem [{\citenamefont {Hawashin}\ \emph {et~al.}(2025)\citenamefont
  {Hawashin}, \citenamefont {Rong},\ and\ \citenamefont
  {Scherer}}]{Hawashin:2024dpp}%
  \BibitemOpen
  \bibfield  {author} {\bibinfo {author} {\bibfnamefont {B.}~\bibnamefont
  {Hawashin}}, \bibinfo {author} {\bibfnamefont {J.}~\bibnamefont {Rong}}, \
  and\ \bibinfo {author} {\bibfnamefont {M.~M.}\ \bibnamefont {Scherer}},\
  }\href {\doibase 10.1103/PhysRevLett.134.041602} {\bibfield  {journal}
  {\bibinfo  {journal} {Phys. Rev. Lett.}\ }\textbf {\bibinfo {volume} {134}},\
  \bibinfo {pages} {041602} (\bibinfo {year} {2025})},\ \Eprint
  {http://arxiv.org/abs/2409.10606} {arXiv:2409.10606 [hep-th]} \BibitemShut
  {NoStop}%
\bibitem [{\citenamefont {Boettcher}\ \emph {et~al.}(2012)\citenamefont
  {Boettcher}, \citenamefont {Pawlowski},\ and\ \citenamefont
  {Diehl}}]{Boettcher:2012cm}%
  \BibitemOpen
  \bibfield  {author} {\bibinfo {author} {\bibfnamefont {I.}~\bibnamefont
  {Boettcher}}, \bibinfo {author} {\bibfnamefont {J.~M.}\ \bibnamefont
  {Pawlowski}}, \ and\ \bibinfo {author} {\bibfnamefont {S.}~\bibnamefont
  {Diehl}},\ }\href {\doibase 10.1016/j.nuclphysbps.2012.06.004} {\bibfield
  {journal} {\bibinfo  {journal} {Nucl. Phys. B Proc. Suppl.}\ }\textbf
  {\bibinfo {volume} {228}},\ \bibinfo {pages} {63} (\bibinfo {year} {2012})},\
  \Eprint {http://arxiv.org/abs/1204.4394} {arXiv:1204.4394
  [cond-mat.quant-gas]} \BibitemShut {NoStop}%
\bibitem [{\citenamefont {Boettcher}\ \emph {et~al.}(2013)\citenamefont
  {Boettcher}, \citenamefont {Diehl}, \citenamefont {Pawlowski},\ and\
  \citenamefont {Wetterich}}]{Boettcher:2012dh}%
  \BibitemOpen
  \bibfield  {author} {\bibinfo {author} {\bibfnamefont {I.}~\bibnamefont
  {Boettcher}}, \bibinfo {author} {\bibfnamefont {S.}~\bibnamefont {Diehl}},
  \bibinfo {author} {\bibfnamefont {J.~M.}\ \bibnamefont {Pawlowski}}, \ and\
  \bibinfo {author} {\bibfnamefont {C.}~\bibnamefont {Wetterich}},\ }\href
  {\doibase 10.1103/PhysRevA.87.023606} {\bibfield  {journal} {\bibinfo
  {journal} {Phys. Rev. A}\ }\textbf {\bibinfo {volume} {87}},\ \bibinfo
  {pages} {023606} (\bibinfo {year} {2013})},\ \Eprint
  {http://arxiv.org/abs/1209.5641} {arXiv:1209.5641 [cond-mat.quant-gas]}
  \BibitemShut {NoStop}%
\bibitem [{\citenamefont {Boettcher}\ \emph {et~al.}(2014)\citenamefont
  {Boettcher}, \citenamefont {Pawlowski},\ and\ \citenamefont
  {Wetterich}}]{Boettcher:2013kia}%
  \BibitemOpen
  \bibfield  {author} {\bibinfo {author} {\bibfnamefont {I.}~\bibnamefont
  {Boettcher}}, \bibinfo {author} {\bibfnamefont {J.~M.}\ \bibnamefont
  {Pawlowski}}, \ and\ \bibinfo {author} {\bibfnamefont {C.}~\bibnamefont
  {Wetterich}},\ }\href {\doibase 10.1103/PhysRevA.89.053630} {\bibfield
  {journal} {\bibinfo  {journal} {Phys.Rev.}\ }\textbf {\bibinfo {volume}
  {A89}},\ \bibinfo {pages} {053630} (\bibinfo {year} {2014})},\ \Eprint
  {http://arxiv.org/abs/1312.0505} {arXiv:1312.0505 [cond-mat.quant-gas]}
  \BibitemShut {NoStop}%
%\%CITATION = ARXIV:1312.0505;\%\%
\bibitem [{\citenamefont {Khan}\ \emph {et~al.}(2015)\citenamefont {Khan},
  \citenamefont {Pawlowski}, \citenamefont {Rennecke},\ and\ \citenamefont
  {Scherer}}]{Khan:2015puu}%
  \BibitemOpen
  \bibfield  {author} {\bibinfo {author} {\bibfnamefont {N.}~\bibnamefont
  {Khan}}, \bibinfo {author} {\bibfnamefont {J.~M.}\ \bibnamefont {Pawlowski}},
  \bibinfo {author} {\bibfnamefont {F.}~\bibnamefont {Rennecke}}, \ and\
  \bibinfo {author} {\bibfnamefont {M.~M.}\ \bibnamefont {Scherer}},\
  }\href@noop {} {\  (\bibinfo {year} {2015})},\ \Eprint
  {http://arxiv.org/abs/1512.03673} {arXiv:1512.03673 [hep-ph]} \BibitemShut
  {NoStop}%
\bibitem [{\citenamefont {Berges}\ \emph {et~al.}(1999)\citenamefont {Berges},
  \citenamefont {Jungnickel},\ and\ \citenamefont {Wetterich}}]{Berges:1997eu}%
  \BibitemOpen
  \bibfield  {author} {\bibinfo {author} {\bibfnamefont {J.}~\bibnamefont
  {Berges}}, \bibinfo {author} {\bibfnamefont {D.~U.}\ \bibnamefont
  {Jungnickel}}, \ and\ \bibinfo {author} {\bibfnamefont {C.}~\bibnamefont
  {Wetterich}},\ }\href {\doibase 10.1103/PhysRevD.59.034010} {\bibfield
  {journal} {\bibinfo  {journal} {Phys. Rev. D}\ }\textbf {\bibinfo {volume}
  {59}},\ \bibinfo {pages} {034010} (\bibinfo {year} {1999})},\ \Eprint
  {http://arxiv.org/abs/hep-ph/9705474} {arXiv:hep-ph/9705474} \BibitemShut
  {NoStop}%
\bibitem [{\citenamefont {Schaefer}\ and\ \citenamefont
  {Wambach}(2005)}]{Schaefer:2004en}%
  \BibitemOpen
  \bibfield  {author} {\bibinfo {author} {\bibfnamefont {B.-J.}\ \bibnamefont
  {Schaefer}}\ and\ \bibinfo {author} {\bibfnamefont {J.}~\bibnamefont
  {Wambach}},\ }\href {\doibase 10.1016/j.nuclphysa.2005.04.012} {\bibfield
  {journal} {\bibinfo  {journal} {Nucl.Phys.}\ }\textbf {\bibinfo {volume}
  {A757}},\ \bibinfo {pages} {479} (\bibinfo {year} {2005})},\ \Eprint
  {http://arxiv.org/abs/nucl-th/0403039} {arXiv:nucl-th/0403039 [nucl-th]}
  \BibitemShut {NoStop}%
%\%CITATION = NUCL-TH/0403039;\%\%
\bibitem [{\citenamefont {Herbst}\ \emph {et~al.}(2013)\citenamefont {Herbst},
  \citenamefont {Pawlowski},\ and\ \citenamefont {Schaefer}}]{Herbst:2013ail}%
  \BibitemOpen
  \bibfield  {author} {\bibinfo {author} {\bibfnamefont {T.~K.}\ \bibnamefont
  {Herbst}}, \bibinfo {author} {\bibfnamefont {J.~M.}\ \bibnamefont
  {Pawlowski}}, \ and\ \bibinfo {author} {\bibfnamefont {B.-J.}\ \bibnamefont
  {Schaefer}},\ }\href {\doibase 10.1103/PhysRevD.88.014007} {\bibfield
  {journal} {\bibinfo  {journal} {Phys. Rev. D}\ }\textbf {\bibinfo {volume}
  {88}},\ \bibinfo {pages} {014007} (\bibinfo {year} {2013})},\ \Eprint
  {http://arxiv.org/abs/1302.1426} {arXiv:1302.1426 [hep-ph]} \BibitemShut
  {NoStop}%
\bibitem [{\citenamefont {Aoki}\ \emph {et~al.}(2014)\citenamefont {Aoki},
  \citenamefont {Sato},\ and\ \citenamefont {Yamada}}]{Aoki:2014ova}%
  \BibitemOpen
  \bibfield  {author} {\bibinfo {author} {\bibfnamefont {K.-I.}\ \bibnamefont
  {Aoki}}, \bibinfo {author} {\bibfnamefont {D.}~\bibnamefont {Sato}}, \ and\
  \bibinfo {author} {\bibfnamefont {M.}~\bibnamefont {Yamada}},\ }\href@noop {}
  {\bibfield  {journal} {\bibinfo  {journal} {Soryushiron Kenkyu}\ }\textbf
  {\bibinfo {volume} {17}},\ \bibinfo {pages} {2} (\bibinfo {year} {2014})},\
  \Eprint {http://arxiv.org/abs/1404.3471} {arXiv:1404.3471 [hep-ph]}
  \BibitemShut {NoStop}%
\bibitem [{\citenamefont {Aoki}\ and\ \citenamefont
  {Yamada}(2015)}]{Aoki:2015hsa}%
  \BibitemOpen
  \bibfield  {author} {\bibinfo {author} {\bibfnamefont {K.-I.}\ \bibnamefont
  {Aoki}}\ and\ \bibinfo {author} {\bibfnamefont {M.}~\bibnamefont {Yamada}},\
  }\href {\doibase 10.1142/S0217751X15501808} {\bibfield  {journal} {\bibinfo
  {journal} {Int. J. Mod. Phys. A}\ }\textbf {\bibinfo {volume} {30}},\
  \bibinfo {pages} {1550180} (\bibinfo {year} {2015})},\ \Eprint
  {http://arxiv.org/abs/1504.00749} {arXiv:1504.00749 [hep-ph]} \BibitemShut
  {NoStop}%
\bibitem [{\citenamefont {Aoki}\ \emph {et~al.}(2018)\citenamefont {Aoki},
  \citenamefont {Kumamoto},\ and\ \citenamefont {Yamada}}]{Aoki:2017rjl}%
  \BibitemOpen
  \bibfield  {author} {\bibinfo {author} {\bibfnamefont {K.-I.}\ \bibnamefont
  {Aoki}}, \bibinfo {author} {\bibfnamefont {S.-I.}\ \bibnamefont {Kumamoto}},
  \ and\ \bibinfo {author} {\bibfnamefont {M.}~\bibnamefont {Yamada}},\ }\href
  {\doibase 10.1016/j.nuclphysb.2018.04.005} {\bibfield  {journal} {\bibinfo
  {journal} {Nucl. Phys. B}\ }\textbf {\bibinfo {volume} {931}},\ \bibinfo
  {pages} {105} (\bibinfo {year} {2018})},\ \Eprint
  {http://arxiv.org/abs/1705.03273} {arXiv:1705.03273 [hep-th]} \BibitemShut
  {NoStop}%
\bibitem [{\citenamefont {Dupuis}\ \emph {et~al.}(2021)\citenamefont {Dupuis},
  \citenamefont {Canet}, \citenamefont {Eichhorn}, \citenamefont {Metzner},
  \citenamefont {Pawlowski}, \citenamefont {Tissier},\ and\ \citenamefont
  {Wschebor}}]{Dupuis:2020fhh}%
  \BibitemOpen
  \bibfield  {author} {\bibinfo {author} {\bibfnamefont {N.}~\bibnamefont
  {Dupuis}}, \bibinfo {author} {\bibfnamefont {L.}~\bibnamefont {Canet}},
  \bibinfo {author} {\bibfnamefont {A.}~\bibnamefont {Eichhorn}}, \bibinfo
  {author} {\bibfnamefont {W.}~\bibnamefont {Metzner}}, \bibinfo {author}
  {\bibfnamefont {J.~M.}\ \bibnamefont {Pawlowski}}, \bibinfo {author}
  {\bibfnamefont {M.}~\bibnamefont {Tissier}}, \ and\ \bibinfo {author}
  {\bibfnamefont {N.}~\bibnamefont {Wschebor}},\ }\href {\doibase
  10.1016/j.physrep.2021.01.001} {\bibfield  {journal} {\bibinfo  {journal}
  {Phys. Rept.}\ }\textbf {\bibinfo {volume} {910}},\ \bibinfo {pages} {1}
  (\bibinfo {year} {2021})},\ \Eprint {http://arxiv.org/abs/2006.04853}
  {arXiv:2006.04853 [cond-mat.stat-mech]} \BibitemShut {NoStop}%
\bibitem [{\citenamefont {Fu}\ \emph {et~al.}(2020)\citenamefont {Fu},
  \citenamefont {Pawlowski},\ and\ \citenamefont {Rennecke}}]{Fu:2019hdw}%
  \BibitemOpen
  \bibfield  {author} {\bibinfo {author} {\bibfnamefont {W.-j.}\ \bibnamefont
  {Fu}}, \bibinfo {author} {\bibfnamefont {J.~M.}\ \bibnamefont {Pawlowski}}, \
  and\ \bibinfo {author} {\bibfnamefont {F.}~\bibnamefont {Rennecke}},\ }\href
  {\doibase 10.1103/PhysRevD.101.054032} {\bibfield  {journal} {\bibinfo
  {journal} {Phys. Rev. D}\ }\textbf {\bibinfo {volume} {101}},\ \bibinfo
  {pages} {054032} (\bibinfo {year} {2020})},\ \Eprint
  {http://arxiv.org/abs/1909.02991} {arXiv:1909.02991 [hep-ph]} \BibitemShut
  {NoStop}%
\bibitem [{\citenamefont {Pisarski}\ and\ \citenamefont
  {Rennecke}(2021)}]{Pisarski:2021qof}%
  \BibitemOpen
  \bibfield  {author} {\bibinfo {author} {\bibfnamefont {R.~D.}\ \bibnamefont
  {Pisarski}}\ and\ \bibinfo {author} {\bibfnamefont {F.}~\bibnamefont
  {Rennecke}},\ }\href {\doibase 10.1103/PhysRevLett.127.152302} {\bibfield
  {journal} {\bibinfo  {journal} {Phys. Rev. Lett.}\ }\textbf {\bibinfo
  {volume} {127}},\ \bibinfo {pages} {152302} (\bibinfo {year} {2021})},\
  \Eprint {http://arxiv.org/abs/2103.06890} {arXiv:2103.06890 [hep-ph]}
  \BibitemShut {NoStop}%
\bibitem [{\citenamefont {Rennecke}\ \emph {et~al.}(2023)\citenamefont
  {Rennecke}, \citenamefont {Pisarski},\ and\ \citenamefont
  {Rischke}}]{Rennecke:2023xhc}%
  \BibitemOpen
  \bibfield  {author} {\bibinfo {author} {\bibfnamefont {F.}~\bibnamefont
  {Rennecke}}, \bibinfo {author} {\bibfnamefont {R.~D.}\ \bibnamefont
  {Pisarski}}, \ and\ \bibinfo {author} {\bibfnamefont {D.~H.}\ \bibnamefont
  {Rischke}},\ }\href {\doibase 10.1103/PhysRevD.107.116011} {\bibfield
  {journal} {\bibinfo  {journal} {Phys. Rev. D}\ }\textbf {\bibinfo {volume}
  {107}},\ \bibinfo {pages} {116011} (\bibinfo {year} {2023})},\ \Eprint
  {http://arxiv.org/abs/2301.11484} {arXiv:2301.11484 [hep-ph]} \BibitemShut
  {NoStop}%
\bibitem [{\citenamefont {Fu}\ \emph {et~al.}(2025)\citenamefont {Fu},
  \citenamefont {Pawlowski}, \citenamefont {Pisarski}, \citenamefont
  {Rennecke}, \citenamefont {Wen},\ and\ \citenamefont {Yin}}]{Fu:2024rto}%
  \BibitemOpen
  \bibfield  {author} {\bibinfo {author} {\bibfnamefont {W.-j.}\ \bibnamefont
  {Fu}}, \bibinfo {author} {\bibfnamefont {J.~M.}\ \bibnamefont {Pawlowski}},
  \bibinfo {author} {\bibfnamefont {R.~D.}\ \bibnamefont {Pisarski}}, \bibinfo
  {author} {\bibfnamefont {F.}~\bibnamefont {Rennecke}}, \bibinfo {author}
  {\bibfnamefont {R.}~\bibnamefont {Wen}}, \ and\ \bibinfo {author}
  {\bibfnamefont {S.}~\bibnamefont {Yin}},\ }\href {\doibase
  10.1103/PhysRevD.111.094026} {\bibfield  {journal} {\bibinfo  {journal}
  {Phys. Rev. D}\ }\textbf {\bibinfo {volume} {111}},\ \bibinfo {pages}
  {094026} (\bibinfo {year} {2025})},\ \Eprint
  {http://arxiv.org/abs/2412.15949} {arXiv:2412.15949 [hep-ph]} \BibitemShut
  {NoStop}%
\bibitem [{\citenamefont {Pawlowski}\ \emph {et~al.}(2025)\citenamefont
  {Pawlowski}, \citenamefont {Rennecke},\ and\ \citenamefont
  {Sattler}}]{Pawlowski:2025jpg}%
  \BibitemOpen
  \bibfield  {author} {\bibinfo {author} {\bibfnamefont {J.~M.}\ \bibnamefont
  {Pawlowski}}, \bibinfo {author} {\bibfnamefont {F.}~\bibnamefont {Rennecke}},
  \ and\ \bibinfo {author} {\bibfnamefont {F.~R.}\ \bibnamefont {Sattler}},\
  }\href@noop {} {\  (\bibinfo {year} {2025})},\ \Eprint
  {http://arxiv.org/abs/2512.20510} {arXiv:2512.20510 [hep-ph]} \BibitemShut
  {NoStop}%
\bibitem [{\citenamefont {Deryagin}\ \emph {et~al.}(1992)\citenamefont
  {Deryagin}, \citenamefont {Grigoriev},\ and\ \citenamefont
  {Rubakov}}]{Deryagin:1992rw}%
  \BibitemOpen
  \bibfield  {author} {\bibinfo {author} {\bibfnamefont {D.~V.}\ \bibnamefont
  {Deryagin}}, \bibinfo {author} {\bibfnamefont {D.~Y.}\ \bibnamefont
  {Grigoriev}}, \ and\ \bibinfo {author} {\bibfnamefont {V.~A.}\ \bibnamefont
  {Rubakov}},\ }\href {\doibase 10.1142/S0217751X92000302} {\bibfield
  {journal} {\bibinfo  {journal} {Int. J. Mod. Phys. A}\ }\textbf {\bibinfo
  {volume} {7}},\ \bibinfo {pages} {659} (\bibinfo {year} {1992})}\BibitemShut
  {NoStop}%
\bibitem [{\citenamefont {Kojo}\ \emph {et~al.}(2010)\citenamefont {Kojo},
  \citenamefont {Hidaka}, \citenamefont {McLerran},\ and\ \citenamefont
  {Pisarski}}]{Kojo:2009ha}%
  \BibitemOpen
  \bibfield  {author} {\bibinfo {author} {\bibfnamefont {T.}~\bibnamefont
  {Kojo}}, \bibinfo {author} {\bibfnamefont {Y.}~\bibnamefont {Hidaka}},
  \bibinfo {author} {\bibfnamefont {L.}~\bibnamefont {McLerran}}, \ and\
  \bibinfo {author} {\bibfnamefont {R.~D.}\ \bibnamefont {Pisarski}},\ }\href
  {\doibase 10.1016/j.nuclphysa.2010.05.053} {\bibfield  {journal} {\bibinfo
  {journal} {Nucl. Phys. A}\ }\textbf {\bibinfo {volume} {843}},\ \bibinfo
  {pages} {37} (\bibinfo {year} {2010})},\ \Eprint
  {http://arxiv.org/abs/0912.3800} {arXiv:0912.3800 [hep-ph]} \BibitemShut
  {NoStop}%
\bibitem [{\citenamefont {Carignano}\ \emph {et~al.}(2010)\citenamefont
  {Carignano}, \citenamefont {Nickel},\ and\ \citenamefont
  {Buballa}}]{Carignano:2010ac}%
  \BibitemOpen
  \bibfield  {author} {\bibinfo {author} {\bibfnamefont {S.}~\bibnamefont
  {Carignano}}, \bibinfo {author} {\bibfnamefont {D.}~\bibnamefont {Nickel}}, \
  and\ \bibinfo {author} {\bibfnamefont {M.}~\bibnamefont {Buballa}},\ }\href
  {\doibase 10.1103/PhysRevD.82.054009} {\bibfield  {journal} {\bibinfo
  {journal} {Phys. Rev. D}\ }\textbf {\bibinfo {volume} {82}},\ \bibinfo
  {pages} {054009} (\bibinfo {year} {2010})},\ \Eprint
  {http://arxiv.org/abs/1007.1397} {arXiv:1007.1397 [hep-ph]} \BibitemShut
  {NoStop}%
\bibitem [{\citenamefont {Fukushima}\ and\ \citenamefont
  {Hatsuda}(2011)}]{Fukushima:2010bq}%
  \BibitemOpen
  \bibfield  {author} {\bibinfo {author} {\bibfnamefont {K.}~\bibnamefont
  {Fukushima}}\ and\ \bibinfo {author} {\bibfnamefont {T.}~\bibnamefont
  {Hatsuda}},\ }\href {\doibase 10.1088/0034-4885/74/1/014001} {\bibfield
  {journal} {\bibinfo  {journal} {Rept.Prog.Phys.}\ }\textbf {\bibinfo {volume}
  {74}},\ \bibinfo {pages} {014001} (\bibinfo {year} {2011})},\ \Eprint
  {http://arxiv.org/abs/1005.4814} {arXiv:1005.4814 [hep-ph]} \BibitemShut
  {NoStop}%
%\%CITATION = ARXIV:1005.4814;\%\%
\bibitem [{\citenamefont {Fukushima}\ and\ \citenamefont
  {Sasaki}(2013)}]{Fukushima:2013rx}%
  \BibitemOpen
  \bibfield  {author} {\bibinfo {author} {\bibfnamefont {K.}~\bibnamefont
  {Fukushima}}\ and\ \bibinfo {author} {\bibfnamefont {C.}~\bibnamefont
  {Sasaki}},\ }\href {\doibase 10.1016/j.ppnp.2013.05.003} {\bibfield
  {journal} {\bibinfo  {journal} {Prog. Part. Nucl. Phys.}\ }\textbf {\bibinfo
  {volume} {72}},\ \bibinfo {pages} {99} (\bibinfo {year} {2013})},\ \Eprint
  {http://arxiv.org/abs/1301.6377} {arXiv:1301.6377 [hep-ph]} \BibitemShut
  {NoStop}%
\bibitem [{\citenamefont {Lee}\ \emph {et~al.}(2015)\citenamefont {Lee},
  \citenamefont {Nakano}, \citenamefont {Tsue}, \citenamefont {Tatsumi},\ and\
  \citenamefont {Friman}}]{Lee:2015bva}%
  \BibitemOpen
  \bibfield  {author} {\bibinfo {author} {\bibfnamefont {T.-G.}\ \bibnamefont
  {Lee}}, \bibinfo {author} {\bibfnamefont {E.}~\bibnamefont {Nakano}},
  \bibinfo {author} {\bibfnamefont {Y.}~\bibnamefont {Tsue}}, \bibinfo {author}
  {\bibfnamefont {T.}~\bibnamefont {Tatsumi}}, \ and\ \bibinfo {author}
  {\bibfnamefont {B.}~\bibnamefont {Friman}},\ }\href {\doibase
  10.1103/PhysRevD.92.034024} {\bibfield  {journal} {\bibinfo  {journal} {Phys.
  Rev. D}\ }\textbf {\bibinfo {volume} {92}},\ \bibinfo {pages} {034024}
  (\bibinfo {year} {2015})},\ \Eprint {http://arxiv.org/abs/1504.03185}
  {arXiv:1504.03185 [hep-ph]} \BibitemShut {NoStop}%
\bibitem [{\citenamefont {Buballa}\ and\ \citenamefont
  {Carignano}(2019)}]{Buballa:2018hux}%
  \BibitemOpen
  \bibfield  {author} {\bibinfo {author} {\bibfnamefont {M.}~\bibnamefont
  {Buballa}}\ and\ \bibinfo {author} {\bibfnamefont {S.}~\bibnamefont
  {Carignano}},\ }\href {\doibase 10.1016/j.physletb.2019.02.045} {\bibfield
  {journal} {\bibinfo  {journal} {Phys. Lett. B}\ }\textbf {\bibinfo {volume}
  {791}},\ \bibinfo {pages} {361} (\bibinfo {year} {2019})},\ \Eprint
  {http://arxiv.org/abs/1809.10066} {arXiv:1809.10066 [hep-ph]} \BibitemShut
  {NoStop}%
\bibitem [{\citenamefont {Pisarski}\ \emph {et~al.}(2021)\citenamefont
  {Pisarski}, \citenamefont {Rennecke}, \citenamefont {Tsvelik},\ and\
  \citenamefont {Valgushev}}]{Pisarski:2020gkx}%
  \BibitemOpen
  \bibfield  {author} {\bibinfo {author} {\bibfnamefont {R.~D.}\ \bibnamefont
  {Pisarski}}, \bibinfo {author} {\bibfnamefont {F.}~\bibnamefont {Rennecke}},
  \bibinfo {author} {\bibfnamefont {A.}~\bibnamefont {Tsvelik}}, \ and\
  \bibinfo {author} {\bibfnamefont {S.}~\bibnamefont {Valgushev}},\ }\href
  {\doibase 10.1016/j.nuclphysa.2020.121910} {\bibfield  {journal} {\bibinfo
  {journal} {Nucl. Phys. A}\ }\textbf {\bibinfo {volume} {1005}},\ \bibinfo
  {pages} {121910} (\bibinfo {year} {2021})},\ \Eprint
  {http://arxiv.org/abs/2005.00045} {arXiv:2005.00045 [nucl-th]} \BibitemShut
  {NoStop}%
\bibitem [{\citenamefont {Pisarski}\ \emph {et~al.}(2020)\citenamefont
  {Pisarski}, \citenamefont {Tsvelik},\ and\ \citenamefont
  {Valgushev}}]{Pisarski:2020dnx}%
  \BibitemOpen
  \bibfield  {author} {\bibinfo {author} {\bibfnamefont {R.~D.}\ \bibnamefont
  {Pisarski}}, \bibinfo {author} {\bibfnamefont {A.~M.}\ \bibnamefont
  {Tsvelik}}, \ and\ \bibinfo {author} {\bibfnamefont {S.}~\bibnamefont
  {Valgushev}},\ }\href {\doibase 10.1103/PhysRevD.102.016015} {\bibfield
  {journal} {\bibinfo  {journal} {Phys. Rev. D}\ }\textbf {\bibinfo {volume}
  {102}},\ \bibinfo {pages} {016015} (\bibinfo {year} {2020})},\ \Eprint
  {http://arxiv.org/abs/2005.10259} {arXiv:2005.10259 [hep-ph]} \BibitemShut
  {NoStop}%
\bibitem [{\citenamefont {Motta}\ \emph {et~al.}(2023)\citenamefont {Motta},
  \citenamefont {Bernhardt}, \citenamefont {Buballa},\ and\ \citenamefont
  {Fischer}}]{Motta:2023pks}%
  \BibitemOpen
  \bibfield  {author} {\bibinfo {author} {\bibfnamefont {T.~F.}\ \bibnamefont
  {Motta}}, \bibinfo {author} {\bibfnamefont {J.}~\bibnamefont {Bernhardt}},
  \bibinfo {author} {\bibfnamefont {M.}~\bibnamefont {Buballa}}, \ and\
  \bibinfo {author} {\bibfnamefont {C.~S.}\ \bibnamefont {Fischer}},\ }\href
  {\doibase 10.1103/PhysRevD.108.114019} {\bibfield  {journal} {\bibinfo
  {journal} {Phys. Rev. D}\ }\textbf {\bibinfo {volume} {108}},\ \bibinfo
  {pages} {114019} (\bibinfo {year} {2023})},\ \Eprint
  {http://arxiv.org/abs/2306.09749} {arXiv:2306.09749 [hep-ph]} \BibitemShut
  {NoStop}%
\bibitem [{\citenamefont {Motta}\ \emph {et~al.}(2024)\citenamefont {Motta},
  \citenamefont {Bernhardt}, \citenamefont {Buballa},\ and\ \citenamefont
  {Fischer}}]{Motta:2024agi}%
  \BibitemOpen
  \bibfield  {author} {\bibinfo {author} {\bibfnamefont {T.~F.}\ \bibnamefont
  {Motta}}, \bibinfo {author} {\bibfnamefont {J.}~\bibnamefont {Bernhardt}},
  \bibinfo {author} {\bibfnamefont {M.}~\bibnamefont {Buballa}}, \ and\
  \bibinfo {author} {\bibfnamefont {C.~S.}\ \bibnamefont {Fischer}},\ }\href
  {\doibase 10.1103/PhysRevD.110.074014} {\bibfield  {journal} {\bibinfo
  {journal} {Phys. Rev. D}\ }\textbf {\bibinfo {volume} {110}},\ \bibinfo
  {pages} {074014} (\bibinfo {year} {2024})},\ \Eprint
  {http://arxiv.org/abs/2406.00205} {arXiv:2406.00205 [hep-ph]} \BibitemShut
  {NoStop}%
\bibitem [{\citenamefont {Motta}\ \emph {et~al.}(2025)\citenamefont {Motta},
  \citenamefont {Bernhardt}, \citenamefont {Buballa},\ and\ \citenamefont
  {Fischer}}]{Motta:2024rvk}%
  \BibitemOpen
  \bibfield  {author} {\bibinfo {author} {\bibfnamefont {T.~F.}\ \bibnamefont
  {Motta}}, \bibinfo {author} {\bibfnamefont {J.}~\bibnamefont {Bernhardt}},
  \bibinfo {author} {\bibfnamefont {M.}~\bibnamefont {Buballa}}, \ and\
  \bibinfo {author} {\bibfnamefont {C.~S.}\ \bibnamefont {Fischer}},\ }\href
  {\doibase 10.1103/PhysRevD.111.074030} {\bibfield  {journal} {\bibinfo
  {journal} {Phys. Rev. D}\ }\textbf {\bibinfo {volume} {111}},\ \bibinfo
  {pages} {074030} (\bibinfo {year} {2025})},\ \Eprint
  {http://arxiv.org/abs/2411.02285} {arXiv:2411.02285 [hep-ph]} \BibitemShut
  {NoStop}%
\bibitem [{\citenamefont {Buballa}\ and\ \citenamefont
  {Carignano}(2015)}]{Buballa:2014tba}%
  \BibitemOpen
  \bibfield  {author} {\bibinfo {author} {\bibfnamefont {M.}~\bibnamefont
  {Buballa}}\ and\ \bibinfo {author} {\bibfnamefont {S.}~\bibnamefont
  {Carignano}},\ }\href {\doibase 10.1016/j.ppnp.2014.11.001} {\bibfield
  {journal} {\bibinfo  {journal} {Prog. Part. Nucl. Phys.}\ }\textbf {\bibinfo
  {volume} {81}},\ \bibinfo {pages} {39} (\bibinfo {year} {2015})},\ \Eprint
  {http://arxiv.org/abs/1406.1367} {arXiv:1406.1367 [hep-ph]} \BibitemShut
  {NoStop}%
\bibitem [{\citenamefont {Goertz}\ \emph {et~al.}(2025)\citenamefont {Goertz},
  \citenamefont {Pastor-Guti{\'e}rrez},\ and\ \citenamefont
  {Pawlowski}}]{Goertz:2024dnz}%
  \BibitemOpen
  \bibfield  {author} {\bibinfo {author} {\bibfnamefont {F.}~\bibnamefont
  {Goertz}}, \bibinfo {author} {\bibfnamefont {{\'A}.}~\bibnamefont
  {Pastor-Guti{\'e}rrez}}, \ and\ \bibinfo {author} {\bibfnamefont {J.~M.}\
  \bibnamefont {Pawlowski}},\ }\href {\doibase 10.1103/7dzj-k6k8} {\bibfield
  {journal} {\bibinfo  {journal} {Phys. Rev. D}\ }\textbf {\bibinfo {volume}
  {112}},\ \bibinfo {pages} {034029} (\bibinfo {year} {2025})},\ \Eprint
  {http://arxiv.org/abs/2412.12254} {arXiv:2412.12254 [hep-th]} \BibitemShut
  {NoStop}%
\bibitem [{\citenamefont {Holthausen}\ \emph {et~al.}(2013)\citenamefont
  {Holthausen}, \citenamefont {Kubo}, \citenamefont {Lim},\ and\ \citenamefont
  {Lindner}}]{Holthausen:2013ota}%
  \BibitemOpen
  \bibfield  {author} {\bibinfo {author} {\bibfnamefont {M.}~\bibnamefont
  {Holthausen}}, \bibinfo {author} {\bibfnamefont {J.}~\bibnamefont {Kubo}},
  \bibinfo {author} {\bibfnamefont {K.~S.}\ \bibnamefont {Lim}}, \ and\
  \bibinfo {author} {\bibfnamefont {M.}~\bibnamefont {Lindner}},\ }\href
  {\doibase 10.1007/JHEP12(2013)076} {\bibfield  {journal} {\bibinfo  {journal}
  {JHEP}\ }\textbf {\bibinfo {volume} {12}},\ \bibinfo {pages} {076} (\bibinfo
  {year} {2013})},\ \Eprint {http://arxiv.org/abs/1310.4423} {arXiv:1310.4423
  [hep-ph]} \BibitemShut {NoStop}%
\bibitem [{\citenamefont {Kubo}\ \emph {et~al.}(2014)\citenamefont {Kubo},
  \citenamefont {Lim},\ and\ \citenamefont {Lindner}}]{Kubo:2014ova}%
  \BibitemOpen
  \bibfield  {author} {\bibinfo {author} {\bibfnamefont {J.}~\bibnamefont
  {Kubo}}, \bibinfo {author} {\bibfnamefont {K.~S.}\ \bibnamefont {Lim}}, \
  and\ \bibinfo {author} {\bibfnamefont {M.}~\bibnamefont {Lindner}},\ }\href
  {\doibase 10.1103/PhysRevLett.113.091604} {\bibfield  {journal} {\bibinfo
  {journal} {Phys. Rev. Lett.}\ }\textbf {\bibinfo {volume} {113}},\ \bibinfo
  {pages} {091604} (\bibinfo {year} {2014})},\ \Eprint
  {http://arxiv.org/abs/1403.4262} {arXiv:1403.4262 [hep-ph]} \BibitemShut
  {NoStop}%
\bibitem [{\citenamefont {Iso}\ \emph {et~al.}(2017)\citenamefont {Iso},
  \citenamefont {Serpico},\ and\ \citenamefont {Shimada}}]{Iso:2017uuu}%
  \BibitemOpen
  \bibfield  {author} {\bibinfo {author} {\bibfnamefont {S.}~\bibnamefont
  {Iso}}, \bibinfo {author} {\bibfnamefont {P.~D.}\ \bibnamefont {Serpico}}, \
  and\ \bibinfo {author} {\bibfnamefont {K.}~\bibnamefont {Shimada}},\ }\href
  {\doibase 10.1103/PhysRevLett.119.141301} {\bibfield  {journal} {\bibinfo
  {journal} {Phys. Rev. Lett.}\ }\textbf {\bibinfo {volume} {119}},\ \bibinfo
  {pages} {141301} (\bibinfo {year} {2017})},\ \Eprint
  {http://arxiv.org/abs/1704.04955} {arXiv:1704.04955 [hep-ph]} \BibitemShut
  {NoStop}%
\bibitem [{\citenamefont {Reichert}\ \emph {et~al.}(2022)\citenamefont
  {Reichert}, \citenamefont {Sannino}, \citenamefont {Wang},\ and\
  \citenamefont {Zhang}}]{Reichert:2021cvs}%
  \BibitemOpen
  \bibfield  {author} {\bibinfo {author} {\bibfnamefont {M.}~\bibnamefont
  {Reichert}}, \bibinfo {author} {\bibfnamefont {F.}~\bibnamefont {Sannino}},
  \bibinfo {author} {\bibfnamefont {Z.-W.}\ \bibnamefont {Wang}}, \ and\
  \bibinfo {author} {\bibfnamefont {C.}~\bibnamefont {Zhang}},\ }\href
  {\doibase 10.1007/JHEP01(2022)003} {\bibfield  {journal} {\bibinfo  {journal}
  {JHEP}\ }\textbf {\bibinfo {volume} {01}},\ \bibinfo {pages} {003} (\bibinfo
  {year} {2022})},\ \Eprint {http://arxiv.org/abs/2109.11552} {arXiv:2109.11552
  [hep-ph]} \BibitemShut {NoStop}%
\bibitem [{\citenamefont {Pasechnik}\ \emph {et~al.}(2024)\citenamefont
  {Pasechnik}, \citenamefont {Reichert}, \citenamefont {Sannino},\ and\
  \citenamefont {Wang}}]{Pasechnik:2023hwv}%
  \BibitemOpen
  \bibfield  {author} {\bibinfo {author} {\bibfnamefont {R.}~\bibnamefont
  {Pasechnik}}, \bibinfo {author} {\bibfnamefont {M.}~\bibnamefont {Reichert}},
  \bibinfo {author} {\bibfnamefont {F.}~\bibnamefont {Sannino}}, \ and\
  \bibinfo {author} {\bibfnamefont {Z.-W.}\ \bibnamefont {Wang}},\ }\href
  {\doibase 10.1007/JHEP02(2024)159} {\bibfield  {journal} {\bibinfo  {journal}
  {JHEP}\ }\textbf {\bibinfo {volume} {02}},\ \bibinfo {pages} {159} (\bibinfo
  {year} {2024})},\ \Eprint {http://arxiv.org/abs/2309.16755} {arXiv:2309.16755
  [hep-ph]} \BibitemShut {NoStop}%
\bibitem [{\citenamefont {Goertz}\ \emph {et~al.}(2023)\citenamefont {Goertz},
  \citenamefont {Pastor-Guti\'errez},\ and\ \citenamefont
  {Pawlowski}}]{Goertz:2023nii}%
  \BibitemOpen
  \bibfield  {author} {\bibinfo {author} {\bibfnamefont {F.}~\bibnamefont
  {Goertz}}, \bibinfo {author} {\bibfnamefont {A.}~\bibnamefont
  {Pastor-Guti\'errez}}, \ and\ \bibinfo {author} {\bibfnamefont {J.~M.}\
  \bibnamefont {Pawlowski}},\ }\href {\doibase 10.1103/PhysRevD.108.095019}
  {\bibfield  {journal} {\bibinfo  {journal} {Phys. Rev. D}\ }\textbf {\bibinfo
  {volume} {108}},\ \bibinfo {pages} {095019} (\bibinfo {year} {2023})},\
  \Eprint {http://arxiv.org/abs/2307.11148} {arXiv:2307.11148 [hep-ph]}
  \BibitemShut {NoStop}%
\bibitem [{\citenamefont {de~Boer}\ \emph {et~al.}(2025)\citenamefont
  {de~Boer}, \citenamefont {Kubo}, \citenamefont {Lindner},\ and\ \citenamefont
  {Reinig}}]{deBoer:2025vpx}%
  \BibitemOpen
  \bibfield  {author} {\bibinfo {author} {\bibfnamefont {T.}~\bibnamefont
  {de~Boer}}, \bibinfo {author} {\bibfnamefont {J.}~\bibnamefont {Kubo}},
  \bibinfo {author} {\bibfnamefont {M.}~\bibnamefont {Lindner}}, \ and\
  \bibinfo {author} {\bibfnamefont {M.}~\bibnamefont {Reinig}},\ }\href@noop {}
  {\  (\bibinfo {year} {2025})},\ \Eprint {http://arxiv.org/abs/2510.12882}
  {arXiv:2510.12882 [hep-ph]} \BibitemShut {NoStop}%
\bibitem [{\citenamefont {Fukushima}\ \emph {et~al.}(2024)\citenamefont
  {Fukushima}, \citenamefont {Hidaka}, \citenamefont {Inoue}, \citenamefont
  {Shigaki},\ and\ \citenamefont {Yamaguchi}}]{Fukushima:2023tpv}%
  \BibitemOpen
  \bibfield  {author} {\bibinfo {author} {\bibfnamefont {K.}~\bibnamefont
  {Fukushima}}, \bibinfo {author} {\bibfnamefont {Y.}~\bibnamefont {Hidaka}},
  \bibinfo {author} {\bibfnamefont {K.}~\bibnamefont {Inoue}}, \bibinfo
  {author} {\bibfnamefont {K.}~\bibnamefont {Shigaki}}, \ and\ \bibinfo
  {author} {\bibfnamefont {Y.}~\bibnamefont {Yamaguchi}},\ }\href {\doibase
  10.1103/PhysRevC.109.L051903} {\bibfield  {journal} {\bibinfo  {journal}
  {Phys. Rev. C}\ }\textbf {\bibinfo {volume} {109}},\ \bibinfo {pages}
  {L051903} (\bibinfo {year} {2024})},\ \Eprint
  {http://arxiv.org/abs/2306.17619} {arXiv:2306.17619 [hep-ph]} \BibitemShut
  {NoStop}%
\bibitem [{\citenamefont {Ihssen}\ \emph
  {et~al.}(2024{\natexlab{a}})\citenamefont {Ihssen}, \citenamefont
  {Pawlowski}, \citenamefont {Sattler},\ and\ \citenamefont
  {Wink}}]{Ihssen:2024miv}%
  \BibitemOpen
  \bibfield  {author} {\bibinfo {author} {\bibfnamefont {F.}~\bibnamefont
  {Ihssen}}, \bibinfo {author} {\bibfnamefont {J.~M.}\ \bibnamefont
  {Pawlowski}}, \bibinfo {author} {\bibfnamefont {F.~R.}\ \bibnamefont
  {Sattler}}, \ and\ \bibinfo {author} {\bibfnamefont {N.}~\bibnamefont
  {Wink}},\ }\href@noop {} {\  (\bibinfo {year} {2024}{\natexlab{a}})},\
  \Eprint {http://arxiv.org/abs/2408.08413} {arXiv:2408.08413 [hep-ph]}
  \BibitemShut {NoStop}%
\bibitem [{\citenamefont {Vafa}\ and\ \citenamefont
  {Witten}(1984)}]{Vafa:1983tf}%
  \BibitemOpen
  \bibfield  {author} {\bibinfo {author} {\bibfnamefont {C.}~\bibnamefont
  {Vafa}}\ and\ \bibinfo {author} {\bibfnamefont {E.}~\bibnamefont {Witten}},\
  }\href {\doibase 10.1016/0550-3213(84)90230-X} {\bibfield  {journal}
  {\bibinfo  {journal} {Nucl. Phys. B}\ }\textbf {\bibinfo {volume} {234}},\
  \bibinfo {pages} {173} (\bibinfo {year} {1984})}\BibitemShut {NoStop}%
\bibitem [{\citenamefont {Ellwanger}\ \emph {et~al.}(1996)\citenamefont
  {Ellwanger}, \citenamefont {Hirsch},\ and\ \citenamefont
  {Weber}}]{Ellwanger:1995qf}%
  \BibitemOpen
  \bibfield  {author} {\bibinfo {author} {\bibfnamefont {U.}~\bibnamefont
  {Ellwanger}}, \bibinfo {author} {\bibfnamefont {M.}~\bibnamefont {Hirsch}}, \
  and\ \bibinfo {author} {\bibfnamefont {A.}~\bibnamefont {Weber}},\ }\href
  {\doibase 10.1007/s002880050073} {\bibfield  {journal} {\bibinfo  {journal}
  {Z.Phys.}\ }\textbf {\bibinfo {volume} {C69}},\ \bibinfo {pages} {687}
  (\bibinfo {year} {1996})},\ \Eprint {http://arxiv.org/abs/hep-th/9506019}
  {arXiv:hep-th/9506019 [hep-th]} \BibitemShut {NoStop}%
%\%CITATION = HEP-TH/9506019;\%\%
\bibitem [{\citenamefont {Gies}(2002)}]{Gies:2002af}%
  \BibitemOpen
  \bibfield  {author} {\bibinfo {author} {\bibfnamefont {H.}~\bibnamefont
  {Gies}},\ }\href {\doibase 10.1103/PhysRevD.66.025006} {\bibfield  {journal}
  {\bibinfo  {journal} {Phys. Rev.}\ }\textbf {\bibinfo {volume} {D66}},\
  \bibinfo {pages} {025006} (\bibinfo {year} {2002})},\ \Eprint
  {http://arxiv.org/abs/hep-th/0202207} {arXiv:hep-th/0202207 [hep-th]}
  \BibitemShut {NoStop}%
%\%CITATION = HEP-TH/0202207;\%\%
\bibitem [{\citenamefont {Pawlowski}\ \emph {et~al.}(2004)\citenamefont
  {Pawlowski}, \citenamefont {Litim}, \citenamefont {Nedelko},\ and\
  \citenamefont {von Smekal}}]{Pawlowski:2003hq}%
  \BibitemOpen
  \bibfield  {author} {\bibinfo {author} {\bibfnamefont {J.~M.}\ \bibnamefont
  {Pawlowski}}, \bibinfo {author} {\bibfnamefont {D.~F.}\ \bibnamefont
  {Litim}}, \bibinfo {author} {\bibfnamefont {S.}~\bibnamefont {Nedelko}}, \
  and\ \bibinfo {author} {\bibfnamefont {L.}~\bibnamefont {von Smekal}},\
  }\href {\doibase 10.1103/PhysRevLett.93.152002} {\bibfield  {journal}
  {\bibinfo  {journal} {Phys.Rev.Lett.}\ }\textbf {\bibinfo {volume} {93}},\
  \bibinfo {pages} {152002} (\bibinfo {year} {2004})},\ \Eprint
  {http://arxiv.org/abs/hep-th/0312324} {arXiv:hep-th/0312324 [hep-th]}
  \BibitemShut {NoStop}%
%\%CITATION = HEP-TH/0312324;\%\%
\bibitem [{\citenamefont {Cyrol}\ \emph {et~al.}(2016)\citenamefont {Cyrol},
  \citenamefont {Fister}, \citenamefont {Mitter}, \citenamefont {Pawlowski},\
  and\ \citenamefont {Strodthoff}}]{Cyrol:2016tym}%
  \BibitemOpen
  \bibfield  {author} {\bibinfo {author} {\bibfnamefont {A.~K.}\ \bibnamefont
  {Cyrol}}, \bibinfo {author} {\bibfnamefont {L.}~\bibnamefont {Fister}},
  \bibinfo {author} {\bibfnamefont {M.}~\bibnamefont {Mitter}}, \bibinfo
  {author} {\bibfnamefont {J.~M.}\ \bibnamefont {Pawlowski}}, \ and\ \bibinfo
  {author} {\bibfnamefont {N.}~\bibnamefont {Strodthoff}},\ }\href {\doibase
  10.1103/PhysRevD.94.054005} {\bibfield  {journal} {\bibinfo  {journal} {Phys.
  Rev.}\ }\textbf {\bibinfo {volume} {D94}},\ \bibinfo {pages} {054005}
  (\bibinfo {year} {2016})},\ \Eprint {http://arxiv.org/abs/1605.01856}
  {arXiv:1605.01856 [hep-ph]} \BibitemShut {NoStop}%
%\%CITATION = ARXIV:1605.01856;\%\%
\bibitem [{\citenamefont {Cyrol}\ \emph
  {et~al.}(2018{\natexlab{a}})\citenamefont {Cyrol}, \citenamefont {Mitter},
  \citenamefont {Pawlowski},\ and\ \citenamefont {Strodthoff}}]{Cyrol:2017qkl}%
  \BibitemOpen
  \bibfield  {author} {\bibinfo {author} {\bibfnamefont {A.~K.}\ \bibnamefont
  {Cyrol}}, \bibinfo {author} {\bibfnamefont {M.}~\bibnamefont {Mitter}},
  \bibinfo {author} {\bibfnamefont {J.~M.}\ \bibnamefont {Pawlowski}}, \ and\
  \bibinfo {author} {\bibfnamefont {N.}~\bibnamefont {Strodthoff}},\ }\href
  {\doibase 10.1103/PhysRevD.97.054015} {\bibfield  {journal} {\bibinfo
  {journal} {Phys. Rev. D}\ }\textbf {\bibinfo {volume} {97}},\ \bibinfo
  {pages} {054015} (\bibinfo {year} {2018}{\natexlab{a}})},\ \Eprint
  {http://arxiv.org/abs/1708.03482} {arXiv:1708.03482 [hep-ph]} \BibitemShut
  {NoStop}%
\bibitem [{\citenamefont {Corell}\ \emph {et~al.}(2018)\citenamefont {Corell},
  \citenamefont {Cyrol}, \citenamefont {Mitter}, \citenamefont {Pawlowski},\
  and\ \citenamefont {Strodthoff}}]{Corell:2018yil}%
  \BibitemOpen
  \bibfield  {author} {\bibinfo {author} {\bibfnamefont {L.}~\bibnamefont
  {Corell}}, \bibinfo {author} {\bibfnamefont {A.~K.}\ \bibnamefont {Cyrol}},
  \bibinfo {author} {\bibfnamefont {M.}~\bibnamefont {Mitter}}, \bibinfo
  {author} {\bibfnamefont {J.~M.}\ \bibnamefont {Pawlowski}}, \ and\ \bibinfo
  {author} {\bibfnamefont {N.}~\bibnamefont {Strodthoff}},\ }\href {\doibase
  10.21468/SciPostPhys.5.6.066} {\bibfield  {journal} {\bibinfo  {journal}
  {SciPost Phys.}\ }\textbf {\bibinfo {volume} {5}},\ \bibinfo {pages} {066}
  (\bibinfo {year} {2018})},\ \Eprint {http://arxiv.org/abs/1803.10092}
  {arXiv:1803.10092 [hep-ph]} \BibitemShut {NoStop}%
\bibitem [{\citenamefont {Ferreira}\ \emph {et~al.}(2025)\citenamefont
  {Ferreira}, \citenamefont {Papavassiliou}, \citenamefont {Pawlowski},\ and\
  \citenamefont {Wink}}]{Ferreira:2025tzo}%
  \BibitemOpen
  \bibfield  {author} {\bibinfo {author} {\bibfnamefont {M.~N.}\ \bibnamefont
  {Ferreira}}, \bibinfo {author} {\bibfnamefont {J.}~\bibnamefont
  {Papavassiliou}}, \bibinfo {author} {\bibfnamefont {J.~M.}\ \bibnamefont
  {Pawlowski}}, \ and\ \bibinfo {author} {\bibfnamefont {N.}~\bibnamefont
  {Wink}},\ }\href {\doibase 10.1140/epjc/s10052-025-15027-7} {\bibfield
  {journal} {\bibinfo  {journal} {Eur. Phys. J. C}\ }\textbf {\bibinfo {volume}
  {85}},\ \bibinfo {pages} {1339} (\bibinfo {year} {2025})},\ \Eprint
  {http://arxiv.org/abs/2508.20568} {arXiv:2508.20568 [hep-ph]} \BibitemShut
  {NoStop}%
\bibitem [{\citenamefont {Gies}\ and\ \citenamefont
  {Wetterich}(2002)}]{Gies:2001nw}%
  \BibitemOpen
  \bibfield  {author} {\bibinfo {author} {\bibfnamefont {H.}~\bibnamefont
  {Gies}}\ and\ \bibinfo {author} {\bibfnamefont {C.}~\bibnamefont
  {Wetterich}},\ }\href {\doibase 10.1103/PhysRevD.65.065001} {\bibfield
  {journal} {\bibinfo  {journal} {Phys.Rev.}\ }\textbf {\bibinfo {volume}
  {D65}},\ \bibinfo {pages} {065001} (\bibinfo {year} {2002})},\ \Eprint
  {http://arxiv.org/abs/hep-th/0107221} {arXiv:hep-th/0107221 [hep-th]}
  \BibitemShut {NoStop}%
%\%CITATION = HEP-TH/0107221;\%\%
\bibitem [{\citenamefont {Gies}\ and\ \citenamefont
  {Wetterich}(2004)}]{Gies:2002hq}%
  \BibitemOpen
  \bibfield  {author} {\bibinfo {author} {\bibfnamefont {H.}~\bibnamefont
  {Gies}}\ and\ \bibinfo {author} {\bibfnamefont {C.}~\bibnamefont
  {Wetterich}},\ }\href {\doibase 10.1103/PhysRevD.69.025001} {\bibfield
  {journal} {\bibinfo  {journal} {Phys.Rev.}\ }\textbf {\bibinfo {volume}
  {D69}},\ \bibinfo {pages} {025001} (\bibinfo {year} {2004})},\ \Eprint
  {http://arxiv.org/abs/hep-th/0209183} {arXiv:hep-th/0209183 [hep-th]}
  \BibitemShut {NoStop}%
%\%CITATION = HEP-TH/0209183;\%\%
\bibitem [{\citenamefont {Braun}(2012)}]{Braun:2011pp}%
  \BibitemOpen
  \bibfield  {author} {\bibinfo {author} {\bibfnamefont {J.}~\bibnamefont
  {Braun}},\ }\href {\doibase 10.1088/0954-3899/39/3/033001} {\bibfield
  {journal} {\bibinfo  {journal} {J.Phys.}\ }\textbf {\bibinfo {volume}
  {G39}},\ \bibinfo {pages} {033001} (\bibinfo {year} {2012})},\ \Eprint
  {http://arxiv.org/abs/1108.4449} {arXiv:1108.4449 [hep-ph]} \BibitemShut
  {NoStop}%
%\%CITATION = ARXIV:1108.4449;\%\%
\bibitem [{\citenamefont {Mitter}\ \emph {et~al.}(2015)\citenamefont {Mitter},
  \citenamefont {Pawlowski},\ and\ \citenamefont
  {Strodthoff}}]{Mitter:2014wpa}%
  \BibitemOpen
  \bibfield  {author} {\bibinfo {author} {\bibfnamefont {M.}~\bibnamefont
  {Mitter}}, \bibinfo {author} {\bibfnamefont {J.~M.}\ \bibnamefont
  {Pawlowski}}, \ and\ \bibinfo {author} {\bibfnamefont {N.}~\bibnamefont
  {Strodthoff}},\ }\href {\doibase 10.1103/PhysRevD.91.054035} {\bibfield
  {journal} {\bibinfo  {journal} {Phys. Rev.}\ }\textbf {\bibinfo {volume}
  {D91}},\ \bibinfo {pages} {054035} (\bibinfo {year} {2015})},\ \Eprint
  {http://arxiv.org/abs/1411.7978} {arXiv:1411.7978 [hep-ph]} \BibitemShut
  {NoStop}%
%\%CITATION = ARXIV:1411.7978;\%\%
\bibitem [{\citenamefont {Cyrol}\ \emph
  {et~al.}(2018{\natexlab{b}})\citenamefont {Cyrol}, \citenamefont {Mitter},
  \citenamefont {Pawlowski},\ and\ \citenamefont {Strodthoff}}]{Cyrol:2017ewj}%
  \BibitemOpen
  \bibfield  {author} {\bibinfo {author} {\bibfnamefont {A.~K.}\ \bibnamefont
  {Cyrol}}, \bibinfo {author} {\bibfnamefont {M.}~\bibnamefont {Mitter}},
  \bibinfo {author} {\bibfnamefont {J.~M.}\ \bibnamefont {Pawlowski}}, \ and\
  \bibinfo {author} {\bibfnamefont {N.}~\bibnamefont {Strodthoff}},\ }\href
  {\doibase 10.1103/PhysRevD.97.054006} {\bibfield  {journal} {\bibinfo
  {journal} {Phys. Rev.}\ }\textbf {\bibinfo {volume} {D97}},\ \bibinfo {pages}
  {054006} (\bibinfo {year} {2018}{\natexlab{b}})},\ \Eprint
  {http://arxiv.org/abs/1706.06326} {arXiv:1706.06326 [hep-ph]} \BibitemShut
  {NoStop}%
%\%CITATION = ARXIV:1706.06326;\%\%
\bibitem [{\citenamefont {Gies}\ and\ \citenamefont
  {Jaeckel}(2006)}]{Gies:2005as}%
  \BibitemOpen
  \bibfield  {author} {\bibinfo {author} {\bibfnamefont {H.}~\bibnamefont
  {Gies}}\ and\ \bibinfo {author} {\bibfnamefont {J.}~\bibnamefont {Jaeckel}},\
  }\href {\doibase 10.1140/epjc/s2006-02475-0} {\bibfield  {journal} {\bibinfo
  {journal} {Eur. Phys. J. C}\ }\textbf {\bibinfo {volume} {46}},\ \bibinfo
  {pages} {433} (\bibinfo {year} {2006})},\ \Eprint
  {http://arxiv.org/abs/hep-ph/0507171} {arXiv:hep-ph/0507171} \BibitemShut
  {NoStop}%
\bibitem [{\citenamefont {Braun}\ and\ \citenamefont
  {Gies}(2006)}]{Braun:2006jd}%
  \BibitemOpen
  \bibfield  {author} {\bibinfo {author} {\bibfnamefont {J.}~\bibnamefont
  {Braun}}\ and\ \bibinfo {author} {\bibfnamefont {H.}~\bibnamefont {Gies}},\
  }\href {\doibase 10.1088/1126-6708/2006/06/024} {\bibfield  {journal}
  {\bibinfo  {journal} {JHEP}\ }\textbf {\bibinfo {volume} {0606}},\ \bibinfo
  {pages} {024} (\bibinfo {year} {2006})},\ \Eprint
  {http://arxiv.org/abs/hep-ph/0602226} {arXiv:hep-ph/0602226 [hep-ph]}
  \BibitemShut {NoStop}%
%\%CITATION = HEP-PH/0602226;\%\%
\bibitem [{\citenamefont {Braun}\ and\ \citenamefont
  {Gies}(2010)}]{Braun:2009ns}%
  \BibitemOpen
  \bibfield  {author} {\bibinfo {author} {\bibfnamefont {J.}~\bibnamefont
  {Braun}}\ and\ \bibinfo {author} {\bibfnamefont {H.}~\bibnamefont {Gies}},\
  }\href {\doibase 10.1007/JHEP05(2010)060} {\bibfield  {journal} {\bibinfo
  {journal} {JHEP}\ }\textbf {\bibinfo {volume} {05}},\ \bibinfo {pages} {060}
  (\bibinfo {year} {2010})},\ \Eprint {http://arxiv.org/abs/0912.4168}
  {arXiv:0912.4168 [hep-ph]} \BibitemShut {NoStop}%
%\%CITATION = 0912.4168;\%\%
\bibitem [{\citenamefont {Braun}\ \emph {et~al.}(2011)\citenamefont {Braun},
  \citenamefont {Fischer},\ and\ \citenamefont {Gies}}]{Braun:2010qs}%
  \BibitemOpen
  \bibfield  {author} {\bibinfo {author} {\bibfnamefont {J.}~\bibnamefont
  {Braun}}, \bibinfo {author} {\bibfnamefont {C.~S.}\ \bibnamefont {Fischer}},
  \ and\ \bibinfo {author} {\bibfnamefont {H.}~\bibnamefont {Gies}},\ }\href
  {\doibase 10.1103/PhysRevD.84.034045} {\bibfield  {journal} {\bibinfo
  {journal} {Phys. Rev. D}\ }\textbf {\bibinfo {volume} {84}},\ \bibinfo
  {pages} {034045} (\bibinfo {year} {2011})},\ \Eprint
  {http://arxiv.org/abs/1012.4279} {arXiv:1012.4279 [hep-ph]} \BibitemShut
  {NoStop}%
\bibitem [{\citenamefont {Pawlowski}(2007)}]{Pawlowski:2005xe}%
  \BibitemOpen
  \bibfield  {author} {\bibinfo {author} {\bibfnamefont {J.~M.}\ \bibnamefont
  {Pawlowski}},\ }\href {\doibase 10.1016/j.aop.2007.01.007} {\bibfield
  {journal} {\bibinfo  {journal} {Annals Phys.}\ }\textbf {\bibinfo {volume}
  {322}},\ \bibinfo {pages} {2831} (\bibinfo {year} {2007})},\ \Eprint
  {http://arxiv.org/abs/hep-th/0512261} {arXiv:hep-th/0512261 [hep-th]}
  \BibitemShut {NoStop}%
%\%CITATION = HEP-TH/0512261;\%\%
\bibitem [{\citenamefont {Goertz}\ and\ \citenamefont
  {Pastor-Guti\'errez}(2025)}]{Goertz:2023pvn}%
  \BibitemOpen
  \bibfield  {author} {\bibinfo {author} {\bibfnamefont {F.}~\bibnamefont
  {Goertz}}\ and\ \bibinfo {author} {\bibfnamefont {A.}~\bibnamefont
  {Pastor-Guti\'errez}},\ }\href {\doibase 10.1140/epjc/s10052-025-13842-6}
  {\bibfield  {journal} {\bibinfo  {journal} {Eur. Phys. J. C}\ }\textbf
  {\bibinfo {volume} {85}},\ \bibinfo {pages} {116} (\bibinfo {year} {2025})},\
  \Eprint {http://arxiv.org/abs/2308.13594} {arXiv:2308.13594 [hep-ph]}
  \BibitemShut {NoStop}%
\bibitem [{\citenamefont {Li}\ \emph {et~al.}(2025)\citenamefont {Li},
  \citenamefont {Pastor-Guti{\'e}rrez}, \citenamefont {Vatani},\ and\
  \citenamefont {Xu}}]{Li:2025tvu}%
  \BibitemOpen
  \bibfield  {author} {\bibinfo {author} {\bibfnamefont {H.-L.}\ \bibnamefont
  {Li}}, \bibinfo {author} {\bibfnamefont {{\'A}.}~\bibnamefont
  {Pastor-Guti{\'e}rrez}}, \bibinfo {author} {\bibfnamefont {S.}~\bibnamefont
  {Vatani}}, \ and\ \bibinfo {author} {\bibfnamefont {L.-X.}\ \bibnamefont
  {Xu}},\ }\href {\doibase 10.1007/JHEP12(2025)020} {\bibfield  {journal}
  {\bibinfo  {journal} {JHEP}\ }\textbf {\bibinfo {volume} {12}},\ \bibinfo
  {pages} {020} (\bibinfo {year} {2025})},\ \Eprint
  {http://arxiv.org/abs/2507.21208} {arXiv:2507.21208 [hep-th]} \BibitemShut
  {NoStop}%
\bibitem [{\citenamefont {Janssen}\ and\ \citenamefont
  {Herbut}(2014)}]{Janssen:2014gea}%
  \BibitemOpen
  \bibfield  {author} {\bibinfo {author} {\bibfnamefont {L.}~\bibnamefont
  {Janssen}}\ and\ \bibinfo {author} {\bibfnamefont {I.~F.}\ \bibnamefont
  {Herbut}},\ }\href {\doibase 10.1103/PhysRevB.89.205403} {\bibfield
  {journal} {\bibinfo  {journal} {Phys. Rev. B}\ }\textbf {\bibinfo {volume}
  {89}},\ \bibinfo {pages} {205403} (\bibinfo {year} {2014})},\ \bibinfo {note}
  {[Addendum: Phys.Rev.B 102, 199902 (2020)]},\ \Eprint
  {http://arxiv.org/abs/1402.6277} {arXiv:1402.6277 [cond-mat.str-el]}
  \BibitemShut {NoStop}%
\bibitem [{\citenamefont {Resch}\ \emph {et~al.}(2019)\citenamefont {Resch},
  \citenamefont {Rennecke},\ and\ \citenamefont {Schaefer}}]{Resch:2017vjs}%
  \BibitemOpen
  \bibfield  {author} {\bibinfo {author} {\bibfnamefont {S.}~\bibnamefont
  {Resch}}, \bibinfo {author} {\bibfnamefont {F.}~\bibnamefont {Rennecke}}, \
  and\ \bibinfo {author} {\bibfnamefont {B.-J.}\ \bibnamefont {Schaefer}},\
  }\href {\doibase 10.1103/PhysRevD.99.076005} {\bibfield  {journal} {\bibinfo
  {journal} {Phys. Rev. D}\ }\textbf {\bibinfo {volume} {99}},\ \bibinfo
  {pages} {076005} (\bibinfo {year} {2019})},\ \Eprint
  {http://arxiv.org/abs/1712.07961} {arXiv:1712.07961 [hep-ph]} \BibitemShut
  {NoStop}%
\bibitem [{\citenamefont {Klinger}\ \emph {et~al.}(2025)\citenamefont
  {Klinger}, \citenamefont {Kaiser},\ and\ \citenamefont
  {Philipsen}}]{Klinger:2025xxb}%
  \BibitemOpen
  \bibfield  {author} {\bibinfo {author} {\bibfnamefont {J.~P.}\ \bibnamefont
  {Klinger}}, \bibinfo {author} {\bibfnamefont {R.}~\bibnamefont {Kaiser}}, \
  and\ \bibinfo {author} {\bibfnamefont {O.}~\bibnamefont {Philipsen}},\ }\href
  {\doibase 10.22323/1.466.0172} {\bibfield  {journal} {\bibinfo  {journal}
  {PoS}\ }\textbf {\bibinfo {volume} {LATTICE2024}},\ \bibinfo {pages} {172}
  (\bibinfo {year} {2025})},\ \Eprint {http://arxiv.org/abs/2501.19251}
  {arXiv:2501.19251 [hep-lat]} \BibitemShut {NoStop}%
\bibitem [{\citenamefont {Cuteri}\ \emph {et~al.}(2021)\citenamefont {Cuteri},
  \citenamefont {Philipsen},\ and\ \citenamefont {Sciarra}}]{Cuteri:2021ikv}%
  \BibitemOpen
  \bibfield  {author} {\bibinfo {author} {\bibfnamefont {F.}~\bibnamefont
  {Cuteri}}, \bibinfo {author} {\bibfnamefont {O.}~\bibnamefont {Philipsen}}, \
  and\ \bibinfo {author} {\bibfnamefont {A.}~\bibnamefont {Sciarra}},\ }\href
  {\doibase 10.1007/JHEP11(2021)141} {\bibfield  {journal} {\bibinfo  {journal}
  {JHEP}\ }\textbf {\bibinfo {volume} {11}},\ \bibinfo {pages} {141} (\bibinfo
  {year} {2021})},\ \Eprint {http://arxiv.org/abs/2107.12739} {arXiv:2107.12739
  [hep-lat]} \BibitemShut {NoStop}%
\bibitem [{\citenamefont {Fej{\H{o}}s}(2014)}]{Fejos:2014qga}%
  \BibitemOpen
  \bibfield  {author} {\bibinfo {author} {\bibfnamefont {G.}~\bibnamefont
  {Fej{\H{o}}s}},\ }\href {\doibase 10.1103/PhysRevD.90.096011} {\bibfield
  {journal} {\bibinfo  {journal} {Phys. Rev. D}\ }\textbf {\bibinfo {volume}
  {90}},\ \bibinfo {pages} {096011} (\bibinfo {year} {2014})},\ \Eprint
  {http://arxiv.org/abs/1409.3695} {arXiv:1409.3695 [hep-ph]} \BibitemShut
  {NoStop}%
\bibitem [{\citenamefont {Fejos}\ and\ \citenamefont
  {Hosaka}(2018)}]{Fejos:2018dyy}%
  \BibitemOpen
  \bibfield  {author} {\bibinfo {author} {\bibfnamefont {G.}~\bibnamefont
  {Fejos}}\ and\ \bibinfo {author} {\bibfnamefont {A.}~\bibnamefont {Hosaka}},\
  }\href {\doibase 10.1103/PhysRevD.98.036009} {\bibfield  {journal} {\bibinfo
  {journal} {Phys. Rev. D}\ }\textbf {\bibinfo {volume} {98}},\ \bibinfo
  {pages} {036009} (\bibinfo {year} {2018})},\ \Eprint
  {http://arxiv.org/abs/1805.08713} {arXiv:1805.08713 [nucl-th]} \BibitemShut
  {NoStop}%
\bibitem [{\citenamefont {Fej{\"o}s}\ and\ \citenamefont
  {Patkos}(2022)}]{Fejos:2021yod}%
  \BibitemOpen
  \bibfield  {author} {\bibinfo {author} {\bibfnamefont {G.}~\bibnamefont
  {Fej{\"o}s}}\ and\ \bibinfo {author} {\bibfnamefont {A.}~\bibnamefont
  {Patkos}},\ }\href {\doibase 10.1103/PhysRevD.105.096007} {\bibfield
  {journal} {\bibinfo  {journal} {Phys. Rev. D}\ }\textbf {\bibinfo {volume}
  {105}},\ \bibinfo {pages} {096007} (\bibinfo {year} {2022})},\ \Eprint
  {http://arxiv.org/abs/2112.14903} {arXiv:2112.14903 [hep-ph]} \BibitemShut
  {NoStop}%
\bibitem [{\citenamefont {Fejos}(2022)}]{Fejos:2022mso}%
  \BibitemOpen
  \bibfield  {author} {\bibinfo {author} {\bibfnamefont {G.}~\bibnamefont
  {Fejos}},\ }\href {\doibase 10.1103/PhysRevD.105.L071506} {\bibfield
  {journal} {\bibinfo  {journal} {Phys. Rev. D}\ }\textbf {\bibinfo {volume}
  {105}},\ \bibinfo {pages} {L071506} (\bibinfo {year} {2022})},\ \Eprint
  {http://arxiv.org/abs/2201.07909} {arXiv:2201.07909 [hep-ph]} \BibitemShut
  {NoStop}%
\bibitem [{\citenamefont {Fejos}\ and\ \citenamefont
  {Hatsuda}(2024)}]{Fejos:2024bgl}%
  \BibitemOpen
  \bibfield  {author} {\bibinfo {author} {\bibfnamefont {G.}~\bibnamefont
  {Fejos}}\ and\ \bibinfo {author} {\bibfnamefont {T.}~\bibnamefont
  {Hatsuda}},\ }\href {\doibase 10.1103/PhysRevD.110.016021} {\bibfield
  {journal} {\bibinfo  {journal} {Phys. Rev. D}\ }\textbf {\bibinfo {volume}
  {110}},\ \bibinfo {pages} {016021} (\bibinfo {year} {2024})},\ \Eprint
  {http://arxiv.org/abs/2404.00554} {arXiv:2404.00554 [hep-ph]} \BibitemShut
  {NoStop}%
\bibitem [{\citenamefont {Braun}\ \emph {et~al.}(2025)\citenamefont {Braun},
  \citenamefont {Chen}, \citenamefont {Fu}, \citenamefont {Gao}, \citenamefont
  {Ihssen}, \citenamefont {Geißel}, \citenamefont {Huang}, \citenamefont
  {Pawlowski}, \citenamefont {Rennecke}, \citenamefont {Sattler}, \citenamefont
  {Schallmo}, \citenamefont {Stoll}, \citenamefont {Tan}, \citenamefont
  {T{\"o}pfel}, \citenamefont {Turnwald}, \citenamefont {Wen}, \citenamefont
  {Wessely}, \citenamefont {Wink}, \citenamefont {Yin},\ and\ \citenamefont
  {Zorbach}}]{fQCD}%
  \BibitemOpen
  \bibfield  {author} {\bibinfo {author} {\bibfnamefont {J.}~\bibnamefont
  {Braun}}, \bibinfo {author} {\bibfnamefont {Y.-r.}\ \bibnamefont {Chen}},
  \bibinfo {author} {\bibfnamefont {W.-j.}\ \bibnamefont {Fu}}, \bibinfo
  {author} {\bibfnamefont {F.}~\bibnamefont {Gao}}, \bibinfo {author}
  {\bibfnamefont {F.}~\bibnamefont {Ihssen}}, \bibinfo {author} {\bibfnamefont
  {A.}~\bibnamefont {Geißel}}, \bibinfo {author} {\bibfnamefont
  {C.}~\bibnamefont {Huang}}, \bibinfo {author} {\bibfnamefont {J.~M.}\
  \bibnamefont {Pawlowski}}, \bibinfo {author} {\bibfnamefont {F.}~\bibnamefont
  {Rennecke}}, \bibinfo {author} {\bibfnamefont {F.~R.}\ \bibnamefont
  {Sattler}}, \bibinfo {author} {\bibfnamefont {B.}~\bibnamefont {Schallmo}},
  \bibinfo {author} {\bibfnamefont {J.}~\bibnamefont {Stoll}}, \bibinfo
  {author} {\bibfnamefont {Y.-y.}\ \bibnamefont {Tan}}, \bibinfo {author}
  {\bibfnamefont {S.}~\bibnamefont {T{\"o}pfel}}, \bibinfo {author}
  {\bibfnamefont {J.}~\bibnamefont {Turnwald}}, \bibinfo {author}
  {\bibfnamefont {R.}~\bibnamefont {Wen}}, \bibinfo {author} {\bibfnamefont
  {J.}~\bibnamefont {Wessely}}, \bibinfo {author} {\bibfnamefont
  {N.}~\bibnamefont {Wink}}, \bibinfo {author} {\bibfnamefont {S.}~\bibnamefont
  {Yin}}, \ and\ \bibinfo {author} {\bibfnamefont {N.}~\bibnamefont
  {Zorbach}},\ }\href@noop {} {\enquote {\bibinfo {title} {{fQCD
  collaboration}},}\ } (\bibinfo {year} {2025}),\ \bibinfo {note}
  {\url{https://fqcd-collaboration.github.io/}}\BibitemShut {NoStop}%
\bibitem [{\citenamefont {Sattler}(2025)}]{Sattler:2025hcg}%
  \BibitemOpen
  \bibfield  {author} {\bibinfo {author} {\bibfnamefont {F.~R.}\ \bibnamefont
  {Sattler}},\ }\emph {\bibinfo {title} {{The Phase Diagram of QCD at High
  Densities}}},\ \href {\doibase 10.11588/heidok.00036639} {Ph.D. thesis},\
  \bibinfo  {school} {U. Heidelberg (main), Heidelberg University} (\bibinfo
  {year} {2025})\BibitemShut {NoStop}%
\bibitem [{\citenamefont {Lo}\ \emph {et~al.}(2013)\citenamefont {Lo},
  \citenamefont {Friman}, \citenamefont {Kaczmarek}, \citenamefont {Redlich},\
  and\ \citenamefont {Sasaki}}]{Lo:2013hla}%
  \BibitemOpen
  \bibfield  {author} {\bibinfo {author} {\bibfnamefont {P.~M.}\ \bibnamefont
  {Lo}}, \bibinfo {author} {\bibfnamefont {B.}~\bibnamefont {Friman}}, \bibinfo
  {author} {\bibfnamefont {O.}~\bibnamefont {Kaczmarek}}, \bibinfo {author}
  {\bibfnamefont {K.}~\bibnamefont {Redlich}}, \ and\ \bibinfo {author}
  {\bibfnamefont {C.}~\bibnamefont {Sasaki}},\ }\href {\doibase
  10.1103/PhysRevD.88.074502} {\bibfield  {journal} {\bibinfo  {journal} {Phys.
  Rev. D}\ }\textbf {\bibinfo {volume} {88}},\ \bibinfo {pages} {074502}
  (\bibinfo {year} {2013})},\ \Eprint {http://arxiv.org/abs/1307.5958}
  {arXiv:1307.5958 [hep-lat]} \BibitemShut {NoStop}%
\bibitem [{\citenamefont {Grossi}\ and\ \citenamefont
  {Wink}(2023)}]{Grossi:2019urj}%
  \BibitemOpen
  \bibfield  {author} {\bibinfo {author} {\bibfnamefont {E.}~\bibnamefont
  {Grossi}}\ and\ \bibinfo {author} {\bibfnamefont {N.}~\bibnamefont {Wink}},\
  }\href {\doibase 10.21468/SciPostPhysCore.6.4.071} {\bibfield  {journal}
  {\bibinfo  {journal} {SciPost Phys. Core}\ }\textbf {\bibinfo {volume} {6}},\
  \bibinfo {pages} {071} (\bibinfo {year} {2023})},\ \Eprint
  {http://arxiv.org/abs/1903.09503} {arXiv:1903.09503 [hep-th]} \BibitemShut
  {NoStop}%
\bibitem [{\citenamefont {Grossi}\ \emph {et~al.}(2021)\citenamefont {Grossi},
  \citenamefont {Ihssen}, \citenamefont {Pawlowski},\ and\ \citenamefont
  {Wink}}]{Grossi:2021ksl}%
  \BibitemOpen
  \bibfield  {author} {\bibinfo {author} {\bibfnamefont {E.}~\bibnamefont
  {Grossi}}, \bibinfo {author} {\bibfnamefont {F.~J.}\ \bibnamefont {Ihssen}},
  \bibinfo {author} {\bibfnamefont {J.~M.}\ \bibnamefont {Pawlowski}}, \ and\
  \bibinfo {author} {\bibfnamefont {N.}~\bibnamefont {Wink}},\ }\href {\doibase
  10.1103/PhysRevD.104.016028} {\bibfield  {journal} {\bibinfo  {journal}
  {Phys. Rev. D}\ }\textbf {\bibinfo {volume} {104}},\ \bibinfo {pages}
  {016028} (\bibinfo {year} {2021})},\ \Eprint
  {http://arxiv.org/abs/2102.01602} {arXiv:2102.01602 [hep-ph]} \BibitemShut
  {NoStop}%
\bibitem [{\citenamefont {Ihssen}\ \emph
  {et~al.}(2024{\natexlab{b}})\citenamefont {Ihssen}, \citenamefont
  {Pawlowski}, \citenamefont {Sattler},\ and\ \citenamefont
  {Wink}}]{Ihssen:2022xkr}%
  \BibitemOpen
  \bibfield  {author} {\bibinfo {author} {\bibfnamefont {F.}~\bibnamefont
  {Ihssen}}, \bibinfo {author} {\bibfnamefont {J.~M.}\ \bibnamefont
  {Pawlowski}}, \bibinfo {author} {\bibfnamefont {F.~R.}\ \bibnamefont
  {Sattler}}, \ and\ \bibinfo {author} {\bibfnamefont {N.}~\bibnamefont
  {Wink}},\ }\href {\doibase 10.1016/j.cpc.2024.109182} {\bibfield  {journal}
  {\bibinfo  {journal} {Comput. Phys. Commun.}\ }\textbf {\bibinfo {volume}
  {300}},\ \bibinfo {pages} {109182} (\bibinfo {year} {2024}{\natexlab{b}})},\
  \Eprint {http://arxiv.org/abs/2207.12266} {arXiv:2207.12266 [hep-th]}
  \BibitemShut {NoStop}%
\bibitem [{\citenamefont {Ihssen}\ \emph {et~al.}(2023)\citenamefont {Ihssen},
  \citenamefont {Sattler},\ and\ \citenamefont {Wink}}]{Ihssen:2023qaq}%
  \BibitemOpen
  \bibfield  {author} {\bibinfo {author} {\bibfnamefont {F.}~\bibnamefont
  {Ihssen}}, \bibinfo {author} {\bibfnamefont {F.~R.}\ \bibnamefont {Sattler}},
  \ and\ \bibinfo {author} {\bibfnamefont {N.}~\bibnamefont {Wink}},\ }\href
  {\doibase 10.1103/PhysRevD.107.114009} {\bibfield  {journal} {\bibinfo
  {journal} {Phys. Rev. D}\ }\textbf {\bibinfo {volume} {107}},\ \bibinfo
  {pages} {114009} (\bibinfo {year} {2023})},\ \Eprint
  {http://arxiv.org/abs/2302.04736} {arXiv:2302.04736 [hep-th]} \BibitemShut
  {NoStop}%
\bibitem [{\citenamefont {Ihssen}\ \emph {et~al.}(2025)\citenamefont {Ihssen},
  \citenamefont {Pawlowski}, \citenamefont {Sattler},\ and\ \citenamefont
  {Wink}}]{Ihssen:2023xlp}%
  \BibitemOpen
  \bibfield  {author} {\bibinfo {author} {\bibfnamefont {F.}~\bibnamefont
  {Ihssen}}, \bibinfo {author} {\bibfnamefont {J.~M.}\ \bibnamefont
  {Pawlowski}}, \bibinfo {author} {\bibfnamefont {F.~R.}\ \bibnamefont
  {Sattler}}, \ and\ \bibinfo {author} {\bibfnamefont {N.}~\bibnamefont
  {Wink}},\ }\href {\doibase 10.1103/PhysRevD.111.036030} {\bibfield  {journal}
  {\bibinfo  {journal} {Phys. Rev. D}\ }\textbf {\bibinfo {volume} {111}},\
  \bibinfo {pages} {036030} (\bibinfo {year} {2025})},\ \Eprint
  {http://arxiv.org/abs/2309.07335} {arXiv:2309.07335 [hep-th]} \BibitemShut
  {NoStop}%
\bibitem [{\citenamefont {Pawlowski}\ \emph {et~al.}(2023)\citenamefont
  {Pawlowski}, \citenamefont {Schneider},\ and\ \citenamefont
  {Wink}}]{Pawlowski:2021tkk}%
  \BibitemOpen
  \bibfield  {author} {\bibinfo {author} {\bibfnamefont {J.~M.}\ \bibnamefont
  {Pawlowski}}, \bibinfo {author} {\bibfnamefont {C.~S.}\ \bibnamefont
  {Schneider}}, \ and\ \bibinfo {author} {\bibfnamefont {N.}~\bibnamefont
  {Wink}},\ }\href {\doibase 10.1016/j.cpc.2023.108711} {\bibfield  {journal}
  {\bibinfo  {journal} {Comput. Phys. Commun.}\ }\textbf {\bibinfo {volume}
  {287}},\ \bibinfo {pages} {108711} (\bibinfo {year} {2023})},\ \Eprint
  {http://arxiv.org/abs/2102.01410} {arXiv:2102.01410 [hep-ph]} \BibitemShut
  {NoStop}%
\bibitem [{\citenamefont {Cyrol}\ \emph {et~al.}(2017)\citenamefont {Cyrol},
  \citenamefont {Mitter},\ and\ \citenamefont {Strodthoff}}]{Cyrol:2016zqb}%
  \BibitemOpen
  \bibfield  {author} {\bibinfo {author} {\bibfnamefont {A.~K.}\ \bibnamefont
  {Cyrol}}, \bibinfo {author} {\bibfnamefont {M.}~\bibnamefont {Mitter}}, \
  and\ \bibinfo {author} {\bibfnamefont {N.}~\bibnamefont {Strodthoff}},\
  }\href {\doibase 10.1016/j.cpc.2017.05.024} {\bibfield  {journal} {\bibinfo
  {journal} {Comput. Phys. Commun.}\ }\textbf {\bibinfo {volume} {C219}},\
  \bibinfo {pages} {346} (\bibinfo {year} {2017})},\ \Eprint
  {http://arxiv.org/abs/1610.09331} {arXiv:1610.09331 [hep-ph]} \BibitemShut
  {NoStop}%
%%CITATION = ARXIV:1610.09331;%%
\bibitem [{\citenamefont {Braun}\ \emph {et~al.}(2026)\citenamefont {Braun},
  \citenamefont {Gei{\ss}el}, \citenamefont {Pawlowski}, \citenamefont
  {Sattler},\ and\ \citenamefont {Wink}}]{Braun:2025gvq}%
  \BibitemOpen
  \bibfield  {author} {\bibinfo {author} {\bibfnamefont {J.}~\bibnamefont
  {Braun}}, \bibinfo {author} {\bibfnamefont {A.}~\bibnamefont {Gei{\ss}el}},
  \bibinfo {author} {\bibfnamefont {J.~M.}\ \bibnamefont {Pawlowski}}, \bibinfo
  {author} {\bibfnamefont {F.~R.}\ \bibnamefont {Sattler}}, \ and\ \bibinfo
  {author} {\bibfnamefont {N.}~\bibnamefont {Wink}},\ }\href {\doibase
  10.1016/j.aop.2025.170250} {\bibfield  {journal} {\bibinfo  {journal} {Annals
  Phys.}\ }\textbf {\bibinfo {volume} {484}},\ \bibinfo {pages} {170250}
  (\bibinfo {year} {2026})},\ \Eprint {http://arxiv.org/abs/2503.05580}
  {arXiv:2503.05580 [hep-th]} \BibitemShut {NoStop}%
\bibitem [{\citenamefont {Sattler}\ and\ \citenamefont
  {Pawlowski}(2024)}]{Sattler:2024ozv}%
  \BibitemOpen
  \bibfield  {author} {\bibinfo {author} {\bibfnamefont {F.~R.}\ \bibnamefont
  {Sattler}}\ and\ \bibinfo {author} {\bibfnamefont {J.~M.}\ \bibnamefont
  {Pawlowski}},\ }\href@noop {} {\  (\bibinfo {year} {2024})},\ \Eprint
  {http://arxiv.org/abs/2412.13043} {arXiv:2412.13043 [hep-ph]} \BibitemShut
  {NoStop}%
\end{thebibliography}%
%%%%%%%%%%%%%%%%%%%%%%%%%%%%%%%%%%%%%%%%

\end{document}